\documentstyle[sprocl2,epsfig]{article}
\def\Journal#1#2#3#4{{#1} {\bf #2}, #3 (#4)}

\def\NPB{{\em Nucl. Phys.} B}

\def\PRD{{\em Phys. Rev.} D}

\def\be{\begin{equation}}
\def\ee{\end{equation}}
\def\bea{\begin{eqnarray}}
\def\eea{\end{eqnarray}}

\begin{document}

\title{'t Hooft and Wilson loop ratios in the  QCD plasma}

\author{P. Giovannangeli, C.P. Korthals Altes }

\address{Centre Physique Theorique au CNRS, Case 907, Luminy, F13288,
Marseille, France}

\author{  }

\address{}


\maketitle\abstracts{
The spatial  't Hooft loop
measuring the electric flux and the spatial Wilsonloop measuring the magnetic flux are analyzed in hot SU(N) gauge theory.  Both display area laws. On one hand the tension of the 't Hooft loop is perturbatively calculable, in the same sense as the pressure. We show that the $O(g^3)$ 
contribution is absent.
The ratio of multi-charged 't Hooft loops have a remarkably simple dependence on
the charge, true up to, but not including, $O(g^4)$.
This dependence follows also from a simple model of free screened colour charges.
On the other  hand the surface tension of the  Wilsonloop is 
non-perturbative. But in a model of screened free monopoles at very 
high temperature
the known  area law follows. The density of these monopoles starts to
contribute to $O(g^6)$ to the pressure. The ratio of the multicharged Wilson loops 
is calculable and identical to that of the 't Hooft loops.}

\section{Introduction}
\label{sec:intro}

The plasma phase of QCD is nowadays object of experimental study at RHIC.

The equilibrium properties of the plasma
 have been analyzed by numerical simulations and
analytic studies. The latter mainly through the use of perturbation theory. The first obvious question is how we can distinguish between a plasma phase and a hadron phase.  

One order parameter is the free energy of a single quark or Wilson line. Another the 
free energy of a single Dirac monopole. The first one is the order 
parameter for a discrete global symmetry, called electric $Z(N)$ symmetry.
 This symmetry is a symmetry of the Euclidean path integral, as long as only $Z(N)$ neutral fields appear. It is 
broken in the high T phase, as signalled by the non vanishing VEV of 
the  Wilson line. The symmetry of the Dirac monopole line is always
 broken.

There are also two canonical order parameters in  QCD : the spatial Wilson
loop and the spatial 't Hooft loop. The first obeys an area law as numerical studies have shown~\cite{karschtension}, {\it {both}} in the confined and in the plasma phase. This corresponds to the Dirac monopole loop having a VEV in both phases. This is in accord with the old idea~\cite{mandelstam,digiacomo} that there is a screening  monopole condensate. 

 The 't Hooft loop has perimeter behaviour in the cold phase (if quarks are absent), and area behaviour in the plasma phase. So the 't Hooft loop is the canonical order parameter that distinguishes the two phases. This corresponds to the free energy of a heavy quark being infinite in the cold and finite in the plasma phase. In the latter phase it   has been calculated analytically~\cite{korthalskovner} and by simulation~\cite{deforcrand}.

What sets the 't Hooft loop and the Wilsonloop apart physically, is that the 't Hooft loop measures the electric colour flux in the plasma, whilst the Wilson loop measures the magnetic colour flux. The plasma is par excellence containing
free or almost free screened colour charges. In such surroundings the area law $\exp{-\sigma(T)A}$ of the 't Hooft loop
follows by a well-known simple physical argument~\cite{teperhart,korthalskovner} to be:
\be
\sigma(T)\sim l_E n_E(T)  
\label{eq:etension}
\ee
with the Debye screening length $l_E$ multiplying the density of electric colour
charge due to the gluons. All what goes into the area law is that gluons are screened, free
and obey a simple thermodynamical distribution law.

Though at extremely high temperatures (extreme
is here $10^{3}$ GeV, so rather academic) simple minded perturbation
theory works for the free energy, it does not work at temperatures 
below. Resummation techniques have been devised to repair the situation~\cite{iancu}.

However to work for the Debye mass
one has to go to even higher temperatures.  So the situation is different from
one observable to another. Though the reason for the bad convergence
is not well understood, the facts are the following: the perturbative 
contributions coming exclusively from the excitations of order $T$ do converge
well. The trouble with the convergence starts when taking into account contributions from
scales comparable to the electric screening mass $O(gT)$ and  the 
magnetic screening mass $O(g^2T)$. For the free energy for example the $O(g^3)$
renders the convergence quite bad. For a systematic view on this, see ref.~\cite{recentkaj} and references therein.

One example where perturbation theory may work better is that of 
the spatial 't Hooft~\cite{thooft} loop, because there $O(g^3)$ are absent as we will show in
this paper. So we may hope that our ratios are not far off its value at the phase transition, and can play a role in the determination of the universality class of the 3d spin model at $T_c$.

In this paper we discuss  in the first place some quantitative
 properties of the electric surface tension 
in the two loop approximation. In particular we show that the surface tension of the 't Hooft loop is identical to the surface tension of 
$Z(N)$ domain walls. The latter are formulated in terms of the potential
for the Wilson line~\cite{bhatta}.

As is well known, the minima of the potential
for the Wilson line are precisely where the Wilsonline takes on centergroup
values. If we denote by $\sigma_k$ the surface tension one gets by tunneling from
the value 1 to the value $\exp{ik{2\pi\over N}}$ through the straight path
connecting them, then we find to one loop order {\it{and}} two loop order:
\be
{\sigma_k\over {k\sigma_1}}={N-k\over{N-1}}
\label{eq:ratio}
\ee

We will show that there is a one to one correspondence between the $\sigma_k$ and the surface tension of an 't Hooft loop with multiple center group
charge $k$.

So  eq.(\ref{eq:ratio})  tells that the 't Hooft loop with the strength of k charges, has 
less energy than k ``fundamental'', or singly charged 't Hooft loops. Thus fundamental 't Hooft loops
will attract each other. The effect 
disappears for large N and finite $k$ as expected in a theory without interactions, i.e. without correlations for loops. For $k$ on the order of $N$ this is not true any longer.

By a trivial rewriting of eq.(\ref{eq:ratio}): 
\be
{\sigma_k\over {\sigma_1}}={k(N-k)\over{N-1}}
\ee
we see that charge conjugation is respected, as  obviously should be the case. This means that only
 from
N=4 on the ratio can take non-trivial values.

When we define the coupling such that the one loop effects disappear 
in the renormalization of the gauge coupling in the reduced model we
have:
\be
\sigma_1={4\pi^2T^2(N-1)\over{3\sqrt{3g^2(T)N}}}
\big(1-(15.2785..-11/3(\gamma_E+1/22)){g^2(T)N\over{(4\pi)^2}}\big)
\label{eq:tension}
\ee
There is a slight modification in comparison with ref.~\cite{bhatta} due to 
a different definition of the gauge coupling there. $\gamma_E=0.577215...$ is Euler's 
constant.

Right at the critical temperature there is the idea of the universality
class of $Z(N)$ spin systems being the same as that of the SU(N) gauge theory
in one dimension higher.

For N=2 or 3 the $Z(N)$ spin system in question is unambiguously defined, and
lattice data have given strong support to this universality.
In particular the SU(2) surface tension has the same critical exponent as its
analogue, the order-order interface, in 3d Z(2) Ising spin systems~\cite{deforcrand}.

But for N larger than 3 the space of couplings defines more than one model
and, what is worse, the critical behaviour of these models is not unique.
For example the family of Potts models has  weights such that for
equal spins the energy is 1, whereas for unequal spins the energy  is $0$.
 One draws immediately the conclusion that for 
{\it{all}}
order-order interfaces the tension is the same.
 So our ratios, if still valid at the transition, would exclude Potts models.
 
 Their  transition
is known to be first order. For clock models it is second order in d=3, 
and infinite order in d=2, for $N\ge 4$.
For Villain models the transition
is infinite order for $N\ge 4$ in d=2, but for d=3 we have not been able
to find any litterature.

On the other hand a recent paper~\cite{wingate} sees a first order transition for the SU(4) pure gauge theory.

The lay-out of this paper is as follows. In section~\ref{sec:elementary} we give the simple intuitive argument that leads to the area law for the
't Hooft loop, eq.(\ref{eq:etension}). 

Section~\ref{sec:quantitative} gives the quantitative formalism, and explains the identity of $Z(N)$ domain wall tension and surface tension of the 
't Hooft loop.

Section~\ref{sec:oneloopratio} analyzes the symmetries of the potential term in the
effective potential and derives the one loop result for the ratio.

Section~\ref{sec:twoloopratio} derives the two loop result for the ratio.

Section~\ref{sec:kinetic} treats the kinetic term in the effective potential in more detail to justify the procedure used in the section before.  It is a little technical, and can be skipped by the reader as the rest of
 the paper can be read independently.
We show that at two loop order the wings of the profile of the "soliton" 
spanned by the 't Hooft loop include the same 
non-perturbative effects as the Debye mass. Nethertheless the integrated
profile, i.e. the surface tension, is finite to two loop order. It ignores to this order
contributions from the electric and magnetic screening scales. This is like in the free energy.
 A new result is that we show why there are no $O(g^3)$ corrections to the tension,
unlike for the free energy.

In section~\ref{sec:absolutemin} we show that the straight path between minima
is certainly a local minimum. For SU(4) numerical analysis shows it is also
a global minimum.

In section~\ref{sec:lesssimple} we revisit the simple method of section~\ref{sec:elementary} and extend it to incorporate the counting rules giving rise
to the ratios and rederive the latter.

In section ~\ref{sec:wilson} we analyze the Wilsonloop in terms of a
speculative monopole model, which has the virtue to predict the same ratios as for the 't Hooft loops.

In the last section we draw conclusions and view the prospects.

To avoid clutter in the formulas we have put the coupling $g=1$. In sections~\ref{sec:oneloopratio}, ~\ref{sec:twoloopratio} and ~\ref{sec:kinetic}  we have put the coupling back in the equations.

\section{Elementary facts}\label{sec:elementary}

Spatial Wilson loops $W(L)$ have been measured since long in simulations.
They are expessed in terms of the vector potential as:
\be
W(L)=Tr{\cal{P}}\exp{i\int_Ld\vec l.\vec A}
\ee

Here $\vec A=\vec A^a\lambda_a$, the Gell-Mann generators being normalized to ${1\over 2}$. The loop, being in the fundamental representation, carries the Z(N) charge $k=1$.

We will be interested in Wilsonloops with multiple $Z(N)$ charge $k$. That is, if $k$ quarks are involved in the representation it has charge $k$ and such a  loop will be written as $W_k(L)$.

Spatial `t Hooft loops $V_l(L)$ have only recently been advocated~\cite{korthalskovner} as useful orderparameters.  The 't Hooft loop is defined as a gauge transformation
that has a jump on a surface with $L$ as border, and the jump $\exp{il{2\pi\over N}}$ is in the 
centergroup of the gauge group $SU(N)$. More explicitely, if $\Omega_l(\vec x)\equiv\exp{i\omega_l(\vec x)}$ is such a gauge transformation in the defining representation of $SU(N)$, then
\be
V_l(L)=\exp{i\vec D\omega_l.\vec E}
\ee
is such an operator.
Our notation is $\vec E\equiv \vec E^a\lambda_a$. The canonical 
commutation relations read:
\be
[E_j^a(\vec x), A_k^b(\vec y)]={1\over i}\delta^{a,b}\delta_{j,k}\delta(\vec x-\vec y)
\ee

This definition suffices for the 't Hooft
operator to have a uniquely determined action on a physical state as shown in the next subsection.
The spatial Wilson loop is of course explicitely gauge invariant.

They obey the 't Hooft commutation relations~\cite{thooft}:
\be
V_l(L)W_k(L')V_l(L)^{-1}=\exp{(ikl{2\pi\over N}n(L,L'))}W_k(L')
\label{eq:fundamental}
\ee

The looping number of the two loops is written as $n(L,L')$. It can take
all integer values.
As is well known this commutation relation imposes that at least one of
the loops must follow an  area law~\cite{thooft}. 

The 't Hooft loop captures electric colour flux, and the Wilson loop 
magnetic colour flux. We will see in this section and in section~\ref{sec:wilson} more precisely how.

In the  subsection we show how the simple formula eq.~(\ref{eq:etension}) comes about.
\subsection{A simple analysis of the 't Hooft loop tension}\label{subsec:simple}

The definition of the spacelike singly charged 't Hooft loop $V_1(L)$ is that of a discontinuous gauge transformation. We can rewrite it as a loop measuring the 
 flux of colour hypercharge:
\be
\label{eq:eflux}
V_1(L)=\exp{i{4\pi\over N}\int_{S(L)} TrY\vec E.d\vec S}
\label{eq:gaugetransformv}
\ee
The canonical electric fieldoperator $\vec E$ is  projected onto the hypercharge matrix \\*
$Y=\hbox{diag}(1,\cdots,1,-N+1)$. 

 This operator, despite the fact that it transforms under gauge transformations as
an adjoint, has a unique and gauge invariant action on physical states, as was expounded in ref.~\cite{korthalskovner2}. 
It is worthwhile to recall this.

 First note that the 't Hooft commutation relation is reproduced by our ``flux'' definition in eq. (\ref{eq:eflux}),
with $l=1$. This follows from the canonical commutation relations between electric
field strength and vector potentials.
Second, physical states can contain non-local gauge invariant quantities
in the form of spatial Wilson loops. The action of $V_1(L)$ on such a loop is fixed by the 't Hooft commutation relation and the looping factor.
So the action on a physical state is the same for both expressions for the loop.

From this it follows there is nothing sacred about the hypercharge direction $Y$. A regular gauge transform of $V_1(L)$:
\be
U(\Omega)V_1(L)U^{\dagger}(\Omega)=\exp{i{4\pi\over N}\int Tr\Omega^{\dagger}Y\Omega\vec E.d\vec S}
\ee

\noindent has the same effect in the 't Hooft commutation relation, so acts the same on a physical state. What is sacred is the normalization in front of
the surface integral, because that determines the  centergoup transformation $\exp{i{2\pi\over N}\Omega Y\Omega^{\dagger}}$, which equals $\exp{i{2\pi\over N}}$ whatever $\Omega$ is.

The thermal average $<V_1(L)>$ has an area law behaviour in the deconfined phase
and it is qualitatively and quantitatively understood why~\cite{korthalskovner}. It is 
simply the consequence of the gluonic colour charges being free, and screened.
To a first approximation screening means that only charges can contribute to the flux, that live in a slab containing the loop of thickness $l_E$, the screening length~\footnote{This is an overestimate.See ref.~\cite{teperhart} for better.} The flux from one single  hypercharge $N$ in the slab will be $\pi$ because only one half of the flux goes through the loop $L$. So the value of the loop from the single charge equals $\exp{i\pi}=-1$. 
To understand this in a slightly different way
consider the adjoint Wilsonloop, given by $TrP\exp{i\int_{L^{\prime}}d\vec l.\vec A_{adj}}$.

 When it loops the 't Hooft loop once
nothing happens because it does not feel the centergroup, and the commutation relation becomes, using eq.(\ref{eq:eflux}) and the canonical commutator of $\vec E$ and $\vec A$:
\be
V_k(L)W_{adj}(L')V_k(L)^{-1}=\exp{(ik{2\pi\over N}Y_{adj})}W_{adj}(L')
\label{eq:adjointcom}
\ee

Here the matrix $Y_{adj}$ is the adjoint representation of the hypercharge
$Y$ defined underneath eq.(\ref{eq:gaugetransformv}). So its $N^2-1$ diagonal matrix elements
consist of the differences of the diagonal elements of $Y$, so are $0$ or $\pm N$, so the phase factor is always one.

 The adjoint Wilson loop sends a flux $N$ through the 't Hooft loop. That is
why in the commutation relation it will just produce a factor
$\exp{i{2\pi\over N}N}=1$. But the point like gluon charge sends 
only a flux ${1\over 2}N$ through the 't Hooft loop, hence the factor $-1$. We will use this 
argument in more detail in section~\ref{sec:lesssimple}.

Since the gluons are free the probability $P(l)$ of l gluons being in the slab
will determine the average of the loop~\footnote{This is a well known quantity called the probability generating function 
$\Pi(z)\equiv\sum_l z^lP(l)$ with $z=-1$.}: 
\be
\label{eq:naive}
<V_1(L)>=\sum_l(-1)^lP(l)
\ee

The probability $P(l)$ follows from the grand canonical Gibbs
 distribution.  Far away from a critical point the distribution is centered around the average $\bar l$ with a finite width $w$ proportional to $\sqrt{ \bar l}$.
For the ideal gas and the high temperature Bose gas it is consistent with a Poisson distribution where $w=\sqrt{ \bar l}$.  Hence we will take the Poisson distribution for simplicity,
keeping in mind the caveat of the width. 
 
So with the probability taken Poissonian
\be
P(l)={1\over{l!}}(\bar l)^l\exp{-\bar l}
\label{eq:poisson}
\ee

\noindent one gets for the
 average of the loop
\be
 <V_1(L)>=\exp{-2\bar l }
\label{eq:average}
\ee

\noindent  by combining eq.(\ref{eq:naive}) and eq. (\ref{eq:poisson}). $\bar l$ is the average 
number of charges in the slab. 
Now that number is  the volume $l_E$ times the
area $A$ of the loop multiplying the density $n$
of gluons and antigluons that have non-zero hypercolour charge $\pm N$:

$$\bar l\sim (N-1)T^3l_E A$$

Together with the known screening length $l_E={1\over{T\sqrt{g^2N/3}}}$ we find for
the tension $\sigma_1$ of a singly charged 't Hooft loop $\sigma_1\sim{ (N-1)T^2\over\sqrt{{g^2N}}}$. 

 More generally there is a
temperature dependent proportionality constant $v(T)$ in the exponent of eq. (\ref{eq:average}) due to the distribution not being Poissonian. This constant will not affect the area behaviour, and it will drop out in ratios of multiply
charged loops.

Suppose we want to know what the tension $\sigma_2$ of an 't Hooft
loop with  charge 2 is. Let's consider to that end two unit charge 
't Hooft loops of equal size and parallel at a distance d. We calculate 
their correlation by the same simple method as above.
If $d>>l_E$ there will be no correlation. However for $d\sim l_E$ a charge in between the two loops will give a $-$ sign to {\it{both}} loops so the correlation between the loops will have a plus sign from the whole volume in between. So the effective volume that does contribute will be only on the outside of the two loops and hence smaller than the total effective volume if the
two loops are far apart. So the surface tension of the correlation will
become smaller, the loops start to attract. 

  One of the aims of this paper
is to  do an analytic calculation of these effects for the 't Hooft loop, and the outcome is qualitatively the same as in eq.(\ref{eq:ratio}).
 Quantitatively the corresponding 
surface tension $\sigma_1$ has been calculated long ago~\cite{bhatta} in perturbation
theory, to two loop order $O(g^2)$.

\section{The quantitative approach}\label{sec:quantitative}

The result for the surface tension of the 't Hooft loop is reminiscent 
of semiclassical physics. In fact it can be computed semiclassically
in a systematic way. The formalism will be explained in this section.
We will, in doing so, also show the relation with  the so-called 
Z(N) domain walls~\cite{bhatta}.

It will turn out that both the profile of a domain wall, and the profile
that describes the surface tension are given by the path ordered Wilsonline
\be
P(A_0(\vec x))\equiv P\exp i\int^{\beta}_0d\tau A_0(\vec
x,\tau)\equiv\lim_{n\rightarrow \infty}\prod^n_{k=1}\exp i\Delta
\tau_kA_0(\vec x,\tau_k)
\ee 

\noindent which we can diagonalize to get the gauge invariant 
phases of the loop.
As is well known its Euclidean average can be rewritten as the free energy of
 a heavy quark in the
fundamental representation~\footnote{
We use the phase here as a mathematical device to compute physical observables like 
the average of the spatial 't Hooft loop. Its physical significance has given rise to a lot of discussion~\cite{smilga}}
.

The Wilson line can wind once, twice or more around the periodic temperature
 direction, and the path integral average of the corresponding traces is then written as $Tr\exp{ik{C\over T}}$, k running from 1 to $N-1$, and the matric $C$ is supposed to be traceless and diagonal $N\hbox{x}N$ .

We define now the effective potential $U(C)$ in a box of size $L_xL_yL_z$
as the constrained path integral:
\be 
\exp{-L_xL_yU(C)}=\int DA \Pi_{z,k}\delta\big(Tr\exp{ik{C\over T}}-Tr\overline{ P^k(A_0)}\big)\exp{-{1\over{g^2}}S(A)}
\label{eq:constrained}
\ee

In this definition  the  area $L_xL_y$
of the $x-y$ cross section of the box factorizes out. 
$k$ is the winding of the Wilson line. The bar stands for averaging over x and y directions. The $z$ dependence in the diagonal matrix $C$ and the 
Wilsonline is not shown.

The size of the box is supposed to be much 
larger 
than the mass gap of the theory.

This definition is manifestly gauge invariant, and is invariant under
a $Z(N)$ transformation of the $A_0$ variable at a given fized timeslice.
What does change under such a transformation is the value of the Polyakov loop in the constraint. So the argument $C$ is shifted by such a transformation, but the value of the effective action stays the same. This invariance is the electric $Z(N)$ invariance. 

$U(C)$ has been evaluated up and
including two loops~\cite{bhatta,korthalsaltes}.
The calculation is done in terms of a saddle point expansion around a
 background field $B$
in the fourth component of the vectorpotential $A_0=B+gQ_0$. $B$ is chosen diagonal and the saddle point equations tell that $B=C$. $C$ has
N diagonal elements given by $C_1,C_2,\cdots,C_N$ and the constraint
 $\sum_i C_i=0$. It lives in the Cartan subspace of the Lie algebra of $SU(N)$.

To this two loop order the potential is of the form:

\bea
U(C)={1\over{2g^2N}}\sum_{i,j}\widehat K(C_{ij})(\partial_zC_{ij})^2+\sum_{ijk}V(C_{ij},C_{ik})
\label{eq:formpotential}
\eea
where $C_{ij}={C_i-C_j\over{2\pi T}}$. Note that these differences are precisely the diagonal elements of the adjoint representation of $C$. So  the dependence of the effective action is solely on the adjoint representation $C_{adj}$ of $C$~\footnote{Not anymore true in the presence of dynamical quarks}.

Often we will refer to all of the kinetic term as $K(C)$.

In this section we will use the effective action to compute the interface tension between any pair of Z(N) vacua, with a charge difference $\Delta k=\pm 1$
or what amounts to the same: the tension in the area law of  the 't Hooft loop\cite{korthalskovner} $V_1(L)$.

\begin{figure} [h]                                                              
\begin{center}                                                                  
\input{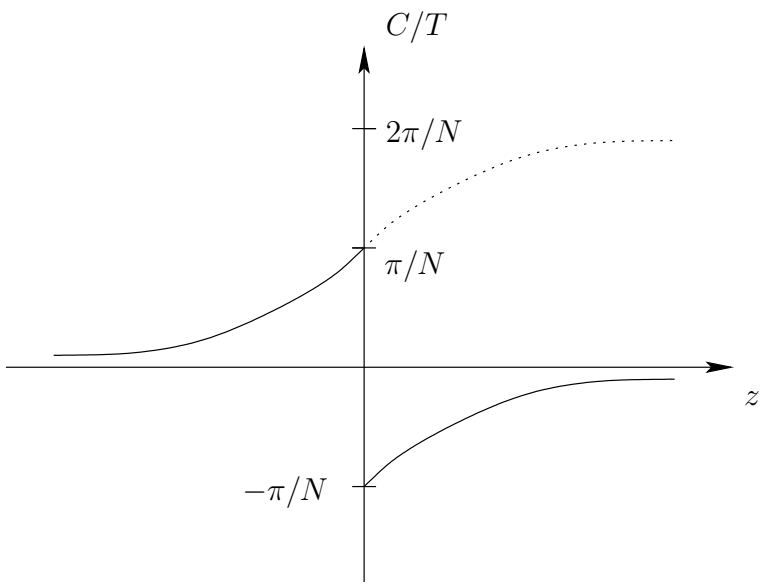}                                                            
\caption{Profile of the Wilson line phase for 't Hooft loop in the x-y plane at $z=0$ (continuous line). This profile minimizes $U(C)$. The broken line is the translate of the right hand branch by  the $\theta$-function in eq.(\ref{eq:boundary}) in the text, and gives the profile one gets by minimizing $U(C)$ between
two Z(N) minima at $C=0$ and $C={2\pi\over N}TY$. }                                

\label{fig:jump}              

\end{center}                                                                    
\end{figure}

The tension of the 't Hooft loop is computed from eq.(\ref{eq:eflux}) by the Gibbs trace:
\be 
TrV_1(L)\exp{-{H\over T}}\sim \exp{-\sigma_1A(L)}
\label{eq:gibbstrace}
\ee
Reexpressing this Gibbs trace in terms of a pathintegral yields precisely the same potential, with boundary conditions $C=0$ and $C={2\pi\over N}TY$ imposed as discontinuity on a surface spanned by the loop $L$ (see fig.(\ref{fig:jump})).

To see that the average of the loop implies naturally the Wilson line effective action, we rewrite the Gibbs trace as a path integral:
\be
<V_1(L)>=\int DA_iDA_0\exp\{-{1\over 2}\int_0^{{1\over T}} dt\int d^3x
\left(\partial_0A_i^a-(D_iA_0)^a-\delta(t)a_i^a\right)^2+(B^a)^2\}
\label{eq:pathintegral}
\ee
where we dropped the normalizing factor.
The source term  $a_i^a$ is the coefficient of the linear term in $\vec E$
in the exponent of $V_1(L)$ in the Gibbs trace. It sits therefore on the surface $S(L)$ of the loop, points in the $Y$ direction of the Cartan algebra, and is orthogonal to the surface $S(L)$ taken to be in the x-y plane.
So it reads on this surface: 

\be
a^a_i({\bf x})=
\delta^{aY}\delta_{iz}{2\pi\over N}\partial_z\theta(z)
\label{eq:boundary}
\ee

It adds to $A_0$ the jump ${2\pi\over N}$, as expected for the
average of the 't Hooft loop! This jump is gauge invariant as we argued below
eq.(\ref{eq:gaugetransformv}). So we have to find the gauge invariant formed out of $A_0$ to meet this invariant boundary condition on $A_0$.
The obvious candidate is $P(A_0)$.

So the road is clear:
 introduce the unit into the pathintegral eq.(\ref{eq:pathintegral}) by the well-known trick:
\be
1=\int DC(z)J(C(z))\Pi_{z,k}\delta\big(Tr\exp{ik{C(z)\over T}}-\overline{TrP^k(A_0)}\big)
\ee

\noindent with $J(C)$ being a Jacobian.
Do  the integrations over the potentials in the presence of the constraint, as in eq.(\ref{eq:constrained}). The pathintegral
 now reduces to an integral over the profiles $C(z)$, with the integrand $J(C(z))\exp{-L_xL_yU(C)}$. 
Letting $L_xL_y$ to infinity tells us to minimize the exponent, with boundary conditions as in fig.(\ref{fig:jump}). 
The phase $C$ jumps at the surface, but its gradient is continuous.
The gradient is the effective electric field appearing in the kinetic energy term of eq.(\ref{eq:formpotential})~\footnote{The resulting profile looks
similar to that of an electrostatic potential due to a dipole layer. There is an obvious question here: can one define a phase for the
Dirac monopole line, and use it as the magnetostatic potential for the profile of the Wilsonloop?}.

Minimization of the exponent $U(C)=K(C)+V(C)$ means that $E(C)\equiv K(C)-V(C)$ is constant in space. Since $K=V=0$ at $z=\pm \infty$ we have  
$E=0$.

\begin{figure} [h]                                                              
\begin{center}                                                                  
\input{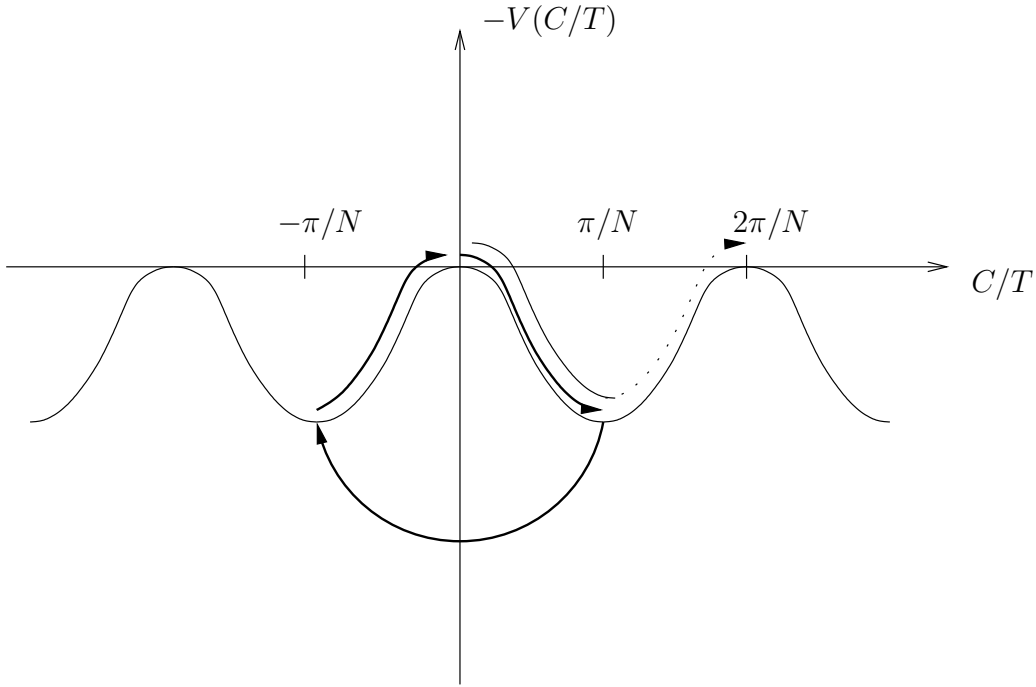}

\caption{Minimizing the effective action $U=K+V$ with two different boundary conditions. The continuous curve corresponds to that of the jump of the
't Hooft loop between ${\pi\over N}$ and $-{\pi\over N}$. The continuous-dotted curve corresponds to the boundary conditions of the domain wall.}

\label{fig:bille}              

\end{center}                                                                    
\end{figure}

 We
have drawn the rolling of the system as a function of $z$ in fig.(\ref{fig:bille}). The continuous curve corresponds to the continuous curve in fig.(\ref{fig:jump}).
So this is the profile for the loop.

In the sections ahead we will compute the surface tension of the 't Hooft
loop as follows. 

Once one knows  $U(C)$ the same tension follows by choosing the appropriate boundary conditions on C (in the case above: $C=0$ and $C={2\pi\over N}TY$) and minimizing the potential. The latter involves finding the path in the Cartan space, that gives an absolute minimum. This will be the method followed 
below and is given in fig.(\ref{fig:jump}) by the combination of continuous and broken line. The broken line follows by shifting the right part of the profile over ${2\pi\over N}$. This undoes the jump at the loop (see eq.(\ref{eq:boundary})). Now the  boundary conditions become $C=0$ at $z=-\infty$ and $C={2\pi\over N}TY$ at $z=\infty$. One recovers the continuous-dotted curve in fig.(\ref{fig:jump}) and fig.(\ref{fig:bille}).

So the minimum value of $U(C)$ is identical with the two types of boundary conditions. This is so because the electric term contains the derivative of the profile, so is the same in both cases. The potential has the
$Z(N)$ periodicity and hence is the same in both cases. In other words:{\it{ surface tensions of $Z(N)$ domain walls and 't Hooft loops are identical}}.

\subsection{Effective potential and Wilsonline correlators}\label{subsec:effectivecorrelators}

The effective potential can be used as well for the calculation of the correlation of Wilsonlines separated
in the z-direction, and averaged over the x-y directions. This can be seen 
easily seen by Laplace transforming our effective action $U(C)$ with a source term $L_xL_yJ.C$. It follows immediately from the definition of $U(C)$, eq.(\ref{eq:constrained}), that we get the free energy $F(J)$, with the source $J$
coupling to the Wilson line.
\be
\exp{-L_xL_y(U(C)-J.C)}=\int DA\exp{\big(-S(A)+J.P(A_0)\big)}\equiv\exp{-L_xL_yF(J)}
\label{eq:correlatorgenerator}
\ee

\noindent with $\partial_CU(C)-J=0$. The dot in $J.C$ means integration over the variable $z$, and in $J.P(A_0)$ integration over all of 3-space.

 This is nothing but the generating functional of 
Wilsonline correlators.

 This is important because it permits us to
connect the asymptotic behaviour of the profile $C$ in the surface tension of the loop to that of the correlator of the imaginary part of two Wilsonlines, which is a gauge invariant and non-perturbative definition  of the Debye mass\cite{arnoldyaffe}.
The Wilsonline correlator behaves according to that definition like $\exp{-{z\over{l_E}}}$, and so will the profile we find by minimization as described above. $l_E$ is the inverse of the Debye mass $m_D$.

It  will permit us  to absorb entirely the $O(g^3)$ term of the surface tension (see section~\ref{sec:kinetic}).

\section{One loop ratio}\label{sec:oneloopratio}

In this section we will compute the surfacetension between two vacua
with a Z(N) charge difference $k$.

From the previous section we know that knowledge of the potential $U(C)$
is not enough to compute the tension. We need the right boundary conditions as well. That is done in the first part of this section. We will search for the domain wall boundary conditions. 

The potential term lives in the space of traceless real diagonal $N\hbox{x}N$
 matrices $C$, i.e. the Cartan subalgebra of SU(N). It has symmetries due to charge conjugation and Z(N) invariance.
Apart from those there is the periodicity in the C-variables (because the potential is defined through the trace  of $\exp{ik{C\over T}}$, $k=1,\cdots,N-1$)  which becomes translation
invariance over $2\pi$ in the space of the $N-1$ variables $C_i$ spanning the Cartan subalgebra.

These invariances lead to the notion of an elementary cell in the Cartan 
subalgebra. It is the smallest simplex in the Cartan subspace that contains all the 
knowledge about the effective potential.

From SU(2) and and SU(3)~\cite{weiss} one can readily generalize what this cell is. It boils down to finding suitable generalizations of the hypercharge Y. There are 
many equivalent definitions but a convenient one is the following~\cite{dvali}:
take the diagonal Cartan matrix $y_1\equiv Y\equiv(1,1,\cdots 1,-N+1)$ and cyclically permute
the diagonal elements to get the Cartan matrices $y_2=(1,\cdots 1,-N+1,1)$, et cetera till
$y_N=(-N+1,1,\cdots,1)$. This set obeys the constraint $\sum_ky_k=0$,
and spans all of the Cartan subalgebra~\footnote{ It is known that the minima of the potential remain fixed in these points $y_k$ 
to all orders of perturbation theory}.

Obviously
\be
\exp{(il{2\pi\over N}y_k)}=\exp{il{2\pi\over N}}
\label{eq:center}
\ee
is a centergroup element, whatever the $y_k$.
We omitted the unit 
matrix on the righthand side. 

The corners of the cell are formed by the diagonal matrices 
\bea
Y_k=\sum_{i=1}^k y_i=(k,\cdots,k,-N+k,\cdots,-N+k)
\label{eq:vertices}
\eea
with $N-k$ entries  $k$ and $k$ entries $-N+k$.

 Because of
eq.(\ref{eq:center}) the corresponding SU(N) group elements 
$\exp{i{2\pi\over N}Y_k}=
\exp{ik{2\pi\over N}}$ are centergroup elements. Within this cell there is a hyperplane that reflects all
$Y_k$ into their counterparts $Y_{N-k}$, i.e. represents charge conjugation invariance of
the potential term with respect to this reflection. Because of Z(N) invariance
there are N of these hyperplanes in the cell reflecting the arbitrariness
of one reference vacuum to another. Only when one breaks the Z(N) invariance
by introducing fundamental multiplets in the theory this plane of charge reflection
is fixed as above. So strictly speaking there is a redundancy in our cell
due to charge conjugation. The reader can find an example of the $SU(4)$
cell in section~\ref{sec:absolutemin}, fig.(\ref{fig:cell}).

Now the explicit form of the one loop potential term~\cite{weiss} is:
\be
V=\int dz{2\over 3}T^4\pi^2\sum_{ij}B_4(C_{ij})
\label{eq:onelooppot}
\ee

\noindent $B_4$ is a Bernoulli polynomial(see appendix A). 

To one loop order we have for the kinetic term simply the classical expression
for the electric field strength:
\be
K=\int dz{1\over {g^2}}Tr(\partial_zC)^2
\ee

Let us now derive our result (\ref{eq:ratio}) for the one loop approximation.
We have to minimize the sum of $K$ and $V$ going from $Y_0$ to $Y_k$. 
We do this under the {\it{assumption}} that the path of minimal action in Cartan space is
the rectilinear one:
\be
C(q)\equiv Y_{0k}(q)={q\over N}Y_k
\label{eq:straightpath}
\ee

\noindent where q lies in the interval $[0,1]$ and parametrizes the path. So on this path $C_{ij(q)}=\pm q$ or $0$.

Let us look at  the one loop potential $V$ along this path.  Now the argument of all Bernoulli
functions  is either $\pm q$  or $0$. $B_4(q)$ is even and vanishes at 
$q=0$. We have to sum over all $i,j$ combinations that give a non zero result for $B_4(q)$
in eq.(\ref{eq:onelooppot}). So it is just a matter of counting.

 From eq.(\ref{eq:vertices}) we see:  if $i <j$ we can pick $i$ in $N-k$ ways and $j$ in $k$ ways to get 
$C_{ij}=q$. Similar for $i >j$.

So along the path $Y_{0k}(q)$ we have 
\be
V(C=Y_{0k}(q))={k(N-k)\over{(N-1)} }V(C=Y_{01}(q))
\label{eq:kdepv}
\ee

\noindent for the one loop potential. 

We now evaluate the kinetic term ${1\over g^2}(\partial_zC)^2$, which can be 
rewritten, using the tracelessness of $C$ as 
${1\over{2g^2N}} \sum_{ij}(\partial_zC_{ij})^2$.

Substitute $C=Y_{0k}(q)$, then again there is no contribution from $C_{ij}$, or it gives the contribution $(\partial_z q)^2$.
Exactly the same counting as for the 
potential terms shows that the sum over $i$ and $j$ gives a dependence
on $k$ identical to that in the potential term: $k(N-k)$. As it is the only 
dependence 
on $k$ we have for the ratio for the effective action  along the path from $0$ to $k$ and and along $0$ to $1$:
\be
U(C=Y_{0k}(q))={k(N-k)\over{N-1}}U(C=Y_{01}(q))
\label{eq:resultpot}
 \ee
Note that the ratio is already true for the potential. A fortiori it will
be true for the tension.

 The tension $\sigma_k$ is obtained by minimizing the effective action 
$U$ along the path $Y_{0k}(q)$. In particular for $k=1$:
\be
U=4\pi^2T^2(N-1)\int dz\big({1\over{g^2N}}(\partial_zq)^2+{T^2\over 3}q^2(1-q)^2\big)
\ee

\noindent with the well known result~\cite{bhatta}:
\be
\sigma_1={4\pi^2T^2(N-1)\over{3\sqrt{3g^2(T)N}}}
\ee
The profiles $q$ as function of $z$ are identical for the multiply charged loops. This follows from minimizing $U(C)$ and using the result eq.(\ref{eq:resultpot}).  This will be the case for two loops as well. Of course we would expect the Debye masses to be the same for any number of loops and
any of the paths $Y_{0k}(q)$.

The counting of contributions leading to the result eq.(\ref{eq:resultpot}) is of course nothing but counting the number of non-zero entries in 
the adjoint representation $Y_{k,adj}(q)$ as underneath eq.(\ref{eq:straightpath}). Up to a sign they are all equal.

We emphasize that we took for granted that the straight path realizes the
absolute minimum. It certainly is a local minimum, and
only in $SU(4)$ we are able by numerical means to establish the truth of
our assumption.

Now we turn to the two loop ratio.

\section{Two loop ratio}\label{sec:twoloopratio}

In this section we show that also the two loop contribution along the straight
path eq.(\ref{eq:straightpath}) to $\sigma_k$  has the dependence $k(N-k)$.

First we analyze the potential term, then the kinetic term.

\subsection{The potential term in two loop order}

We start by recalling the result
for the two loop approximation to the effective potential $V$.

The two loop result is given by the sum of two contributions~\cite{korthalsaltes}~\cite{bhatta} :
\bea
\label{eq:twolooppot}
V_f &=& g^2\sum_{b,c,a}|f^{bca}|^2(\widehat B_2(C_c)\widehat B_2(C_a)-
\widehat B^2_2(0))\nonumber\\
V_P &=&4g^2N\left[\sum_{i<j}\widehat B_3(C_{ij})\widehat B_1(C_{ij})\right]
\eea

The $\widehat B$-functions in this expression are 
the Bernoulli polynomials  given in the appendix.

$V_f$ is the contribution from the free energy graphs, and the constraint
in the definition eq.(\ref{eq:constrained}) tells us there is a renormalization
of the Wilson line phase $4g^2N\widehat B_1$ which adds through the 
lowest order $V$ in eq.(\ref{eq:onelooppot}), or rather its derivative $\widehat B_3$ the quantity  $V_P$ to the effective action. This is useful to keep in mind when we analyze the effective electrostatic action $\cal{L}_E$ in section~\ref{sec:kinetic}.

Let us first look at the contribution of this renormalization of the
Wilson line. It has {\it{per se}} the same form as the lowest order contribution
namely a sum over the index pair $ij$. So along the straight path $Y_{0k}$
this summation gives, as in the lowest order case a factor $k(N-k)$ and
no other k dependence.

The contribution of the two loop free energy graphs needs a little more
analysis. 

What we have to do is to write out the adjoint indices $a,b,c$ on the f-symbol in terms of the 
index pair symbols $ij,jl,li$ from the Cartan basis. In that basis $\lambda_d$ is a set of $N-1$  diagonal and orthogonal matrices with norm ${1\over 2}$ and the remaining $N(N-1)$
$\lambda^{ij}_{kl}={1\over{\sqrt 2}}\delta_{ik}\delta_{jl}$ .  
This is of course nothing else than computing with graphs in the double 
line representation.

We rewrite in this basis the contribution as:

\bea
V_f &=& g^2\sum_{i\neq j}\big(\widehat B_2^2(C_{ij})+2\widehat B_2(C_{ij})\widehat B_2(0)\big)\nonumber \\
   & &+{g^2\over 2}\big(\sum_{i<j<l}+\sum_{l<j<i}\big)\big(\widehat B_2(C_{ij})\widehat B_2(C_{jl})+ 5 \hbox{permutations}\big)
\label{eq:potentialtermtwoloop}
\eea

The first term gives the effect of graphs with one diagonal gluon. It
has the same structure as the one loop, and hence the now familiar
k dependence.

The second and third term differ by the order of the indices
$i,j~\hbox {and}~l$ in the summation. Charge conjugation changes $\lambda^{ij}$ into $\lambda^{ji}$
and so the first term is the charge conjugate of the second.
The arguments of the Bernoulli functions only change sign but Bernoulli  
functions of even order are even under sign change of their argument.
So along the straight path the contribution of the first term is proportional
to $(N-k)k(k-1)$, because the index $i$ must be in the first $N-k$ entries of 
$Y_{0k}(q)=\hbox{diag}(k,\cdots,k,-N+k,\cdots,-N+k)$, the index $j$ in the 
$k$ last entries of $Y_{0k}(q)$, so for $l$ there remain only $k-1$ entries,
because of the ordering.

The second term is according to the same reasoning proportional
to  $k(N-k)(N-k-1)$.

So adding the first two terms give  a proportionality constant $k(N-k)(N-2)$.
Adding all three terms of the potential term  gives simply:
\be
V=k(N-k){4\over 3}\pi^2T^4B_4(q)(1-5{g^2N\over{(4\pi)^2}})
\label{eq:potentialresult}
\ee
and shows that along the straight path the potential term is just 
multiplicatively renormalized by the two loop terms. This generalizes what was noticed  long
ago~\cite{bhatta} for the straight path (the "q-valley") from the $k=0$ to $k=1$ minimum, $Y_{01}(q)$.

So far for the potential term.

\subsection{The kinetic term to two loop order}

The kinetic term one gets by doing a gradient expansion~\cite{bhatta}.
It reads:
\be
K(C)={1\over{2g^2N}}\sum_{ij}\big(1+{g^2N\over {(4\pi)^2}}{11\over 3}(\psi(C_{ij})+
\psi(1-C_{ij})+ \gamma_E+{1\over{22}})\big)(\partial_zC_{ij})^2
\label{eq:kineticterm}
\ee

The sum excludes $i=j$. The $\psi$ are the logarithmic derivatives of Eulers $\Gamma$
function, and develop a pole $-{1\over{C_{ij}}}$ for small $C_{ij}$. They are defined
on the interval $[0,1]$ and periodically extended. So the sum of the two $\psi$
functions is even under $C_{ij}\leftrightarrow -C_{ij}$. Note that for small
profile $C_{ij}$ is {\it{negative}}.

For the moment we will ignore this unphysical behaviour for small $C_{ij}$, and postpone its
discussion till the section~\ref{sec:kinetic}.
We will simply assume in this subsection that 
\be
\psi(C_{ij})(\partial_z C_{ij})^2=0 ~\hbox{if{} i,j such that }{} C_{ij}=0
\label{eq:assume}
\ee
and justify that in the next section.

Then, again it is the now familiar counting: along the straight path $Y_{0k}(q)$ we find easily that the kinetic term 
behaves like:
\be
K(C=Y_{0k})={k(N-k)\over{N-1}}K(C=Y_{01})
\label{eq:kineticresult}
\ee

Our final task is to minimize the full effective potential U, the sum of
the kinetic term eq.(\ref{eq:kineticresult}) and the potential term 
eq.(\ref{eq:potentialresult}) over all possible profiles $q(z)$ along the straight path.

This is standard procedure. We take the full potential and write it as a pure
square and a rest:
\be
U=\int dz\big( K^{{1\over 2}}-V^{{1\over 2}})^2+2\int dz (KV)^{{1\over 2}}
\label{eq:standardmin}
\ee

The rest term is independent of whatever profile one takes, since 
$dz K^{{1\over 2}}={1\over{2g^2N}}\widehat K(q)^{1\over 2}dq$, so the rest integral is a definite integral.

So the minimum of $U$ is reached when 
\be
K^{{1\over 2}}=V^{{1\over 2}}{}\hbox{ for all}{~} z
\label{eq:motion}
\ee

The rest integral then gives the $k$ dependence of the tension as quoted above in eq.(\ref{eq:ratio}).
It is important that in the rest integral the behaviour of $V$ at $q=0$ cancels
out the pole in $K$.

 We have gone into some detail because of an important issue:
the minimization goes wrong for 
$q=O(g^2)$, since
$K$ becomes then 
negative. When computed correctly this does not happen as explained
in the next section. Instead of $O(g^2)$ the correction to the kinetic energy will turn out to be $O(g)$. 

\section{The kinetic energy at small values of the profile}\label{sec:kinetic}

This section gives the justification of our procedure in the previous section.
This justification is based on the effective action approach~\cite{pisarski}. We found it very convenient to use this effective action approach
for the effective potential of the Wilsonline, the constrained path integral eq.(\ref{eq:constrained}).

As a byproduct we compute the well-known first leading correction to the Debye mass~\cite{rebhan}. It shows the expediency of the method.
We also show  how the $O(g^3)$ correction to the tension in eq.(\ref{eq:tension}) does vanish. The latter is a new result.

  So the next order that may contribute is $O(g^4)$. Recent lattice data~\cite{deforcrand} confirm that the two 
loop approximation for the surface tension
works quite well for reasonably low temperatures. Of course only computation
of the $O(g^4)$ term will decide, and this is presently being done~\cite{giovan}.

So let us now come to a quantitative analysis of the kinetic term at small values of the profile.

The kinetic term, eq.(\ref{eq:kineticterm}), has been computed in a gradient expansion
 ${|\vec k|\over C}$. Since $|\vec k|=m_D$, the Debye mass, the 
expansion becomes questionable when $C\sim gT$ or smaller. In fact, simple
powercounting~\cite{linde} tells that for $C=O(g^2)$
 the naive expansion breaks down completely already at one loop, and this is reflected
by the pole $\sim g^2{1\over C}$~\footnote{It was analyzed some time ago how to remedy the kinetic term\cite{bronoff}.}.

\subsection{The electric QCD action with a gauge invariant infrared cut-off}

What we will find by integrating out the hard modes in eq.(\ref{eq:constrained}) is  an adaptation of a well known approach \cite{braatennieto}.

The adaption is precisely this: in the phase $C$ of the Wilsonline we have 
a gauge invariant infrared cut-off of the propagators.

So we start from eq.(\ref{eq:constrained}), pick a background $\xi$ gauge, and 
integrate out only the hard modes.

 To lowest order one gets the following result:
\be
{\cal{L}}_E=Tr(\vec D{\tilde Q_0})^2+\sum_{m,n}TrF_{mn}^2+ {1\over{\xi}}Tr(\vec{\partial}.\vec{\tilde Q}+i\xi [C,{\tilde Q_0}])^2 + V(\tilde Q_0,\vec{ \tilde Q};C)+\tilde U(C)
\label{eq:elagrangian}
\ee
All fields are static and scaled by $\sqrt T$: $\tilde Q_{\mu}={Q_{\mu}(n=0)\over{\sqrt{T}}}$. $\vec D=\vec\partial+ig_3[\vec{\tilde Q},$, $F_{mn}$ is the three dimensional gaugefield strength in terms of $\vec Q$, and $g_3=g\sqrt T$.  The gauge condition is the static remnant of the original 4d $\xi$ background gauge. $\tilde U(C)$ is the result we get for $U(C)$
incorporating only the hard modes\footnote{This term is the analogue of the unit operator term of ref.\cite{braatennieto}}. To the order we are interested in the potential term is:
\be
V(\tilde Q_0,\vec{\tilde Q};C)=Tr m_E^2(C)(\tilde Q_0)^2+ Tr(i[C,\vec{\tilde Q}])^2+4iTr\partial_zC[\tilde Q_z,\tilde Q_0]
\ee

The higher order vertices are of no concern for us in the calculation of
the kinetic energy for small profile.
The potential term  has the form of  the familiar electric QCD action~\cite{braatennieto}, with some 
important differences however. The propagators of the charged gauge particles $\vec{\tilde Q}^{ij}$ in this
3d theory still have a mass proportional to $C_{ij}$. Note that this mass is gauge invariant and hence assures of a gauge invariant infrared regulator. The Higgs mass matrix  $m_E^2$ becomes only in the limit of small $C$ equal to the ${N\over 3}g^2T^2$.  There are also additional couplings
due to the presence of the gauge invariant constraint. These are necessary for the masses
to be gauge invariant.

What we need to know is the mass matrix $m_E^2(C)$ for finite values of $C$.
It has three contributions. For sake of simplicity we will 
limit the discussion to the $SU(2)$ case, where the matrix $C$ is parametrized as $C=\hbox{diag}(\pi Tq,-\pi Tq)$.

The first one is the familiar one that one gets from summing the hard modes
with vertices and propagators coming obtained from the action with background field $C$:
\be
m^2_{3,12}(q)={2\over 3}g^2T^2(1+a_{3,12}q^2)
\label{eq:nocon}
\ee

\noindent for the diagonal ($m_3$) and off-diagonal ($m_{12}$) masses.
What is important is analyticity in $q^2$, because we summed only over the hard modes.

The second and third are due to the presence of the constraint.
In what follows we explain how they come about.
We Fourier analyze the constraint with  variables $\lambda(z)$.
That adds in the exponent in eq. (\ref{eq:constrained}) in addition to the action $-{1\over {g^2}}S$:
\be
i\int dz {\lambda(z)\over{g^2}}.\big({1\over 2}\overline{TrP(A_0)}(z)-{1\over 2}Tr\exp{i{C\over T}}\big)
\label{eq:developcon}
\ee

We have to perform a steepest descent in both $\lambda=\lambda_{cl}+g\lambda_{qu}$ and the field variables $A_0=B+gQ_0$ and $\vec A=g\vec Q$. After expansion of the constraint and the action one finds that the saddle point is

\bea
{i\lambda_{cl}\over{TL_{tr}^2}}&=&{4\pi\partial_z^2q\over{\sin{\pi q}}}\\
\label{eq:saddle}                              
B&=&C
\eea

When we develop the constraint one finds:
\bea
{1\over 2}\overline{TrP(A_0)}-{1\over 2}Tr\exp{i{C\over T}}&=&-{g\over 2}\sin(\pi q){\overline{ Q^3_0(p^0=0)}\over T}\nonumber\\
                                     &-&{g^2\over {8T^2}}\overline{ (Q^3_0(p^0=0) )^2}\cos(\pi q)\nonumber\\
                                     &-&{g^2\over{2T}} \sum_{p^0}{\overline{Q_0^{12}(p^0)Q_0^{21}(-p^0)}\over{(p^0+2\pi qT)}}\sin(\pi q)+O(Q^3)
\label{eq:expansion}
\eea

The Matsubara frequency is $p_0=2\pi T n_0$, $ n_0$ integer.
For notational reasons we have suppressed the $z$ dependence in all the fields. Note there are no heavy diagonal modes, and that the coefficients of the constant modes are different when $q\neq 0$.

So there will be a contribution to $m_E(q)^2$ from the constraint alone by combining the value for $\lambda_{cl}$ in eq. (\ref{eq:saddle}) and the expansion of the constraint. In the limit of small $q$ it reads for the masses $m^2_3(q)$ and $m_{12}^2(q)$ of respectively diagonal and off-diagonal fields $\tilde Q_0$ the same:
\be
m_c^2(q)={\partial_z^2q\over{ q}}
\label{eq:masscon}
\ee

\noindent and for general $q$:
\bea
m^2_{c,3}(q)&=&{\pi\partial_z^2q\over{\sin(\pi q)}}\cos(\pi q)\\
m^2_{c,12}(q)&=&{\partial_z^2q\over{ q}}
\label{eq:massmatrix}
\eea

The third contribution comes from the integration over the $\lambda_{qu}$
variables. 

Their effect is to couple the terms quadratic and higher in $Q_0$ from  the constraint in eq. (\ref{eq:expansion}) to the conventional vertices present in 
the action $S$.

\begin{figure} [h]
\begin{center}
\input{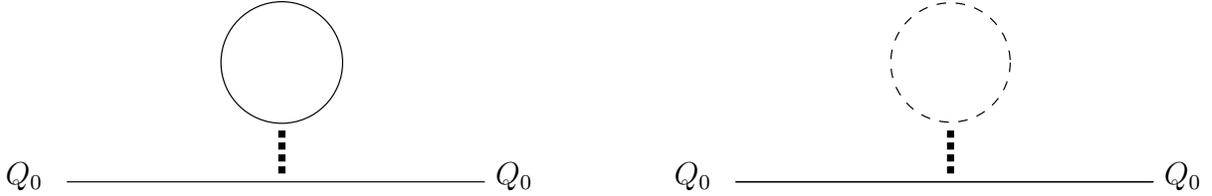}
\caption{The two diagrams with the mixed constraint/conventional vertices, giving the mass $m_{cc}^2$ in eq. (\ref{eq:mixed}). In the loops are only hard modes}
\label{fig:mass}
\end{center}
\end{figure}

For the mass term $Tr\tilde Q_0^2$ these couplings are shown in fig.(\ref{fig:mass})~\footnote{These mixed couplings are important for the consistency of the approach as
the following example may illustrate.

There is a contribution from the mixed vertices with the static $Q_0$ replaced by the heavy $Q_0$ modes with $p^0\neq 0$, the last term  in eq. (\ref{eq:expansion}). When integrated over the heavy modes this 
gives the diagrams in fig. (\ref{fig:mass}) but now with the two lower 
lines  connected to form a loop as well. The upper loops will give again the
derivative of the one loop potential $\sim \widehat{{\tilde B}}_3(q)$ and the
loop with the heavy $Q_0$ modes gives $\widehat{{\tilde B}}_1(q)$. The latter factor carries the gauge dependence necessary to kill the gauge dependence in the two loop free energy graphs, like in eq.(\ref{eq:twolooppot}).This mechanism was first noted by Belyaev~\cite{belyaev}.} .

 The loops contain
the hard modes only and give rise to the derivative of the result for the one loop potential, eq.(\ref{eq:onelooppot}), but with the cubic term omitted (it is due to the zero Matsubara frequency).
As a result we have for the mass of the diagonal mode:
\be
m^2_{cc,3}=-{2\over 3}g^2T^2{\pi q\over {\sin(\pi q)}}\cos(\pi q) (1+2q^2)
\label{eq:mixed}
\ee 
and for the off-diagonal mode:
\be
m^2_{cc,12}=-{2\over 3}g^2T^2 (1+2q^2)
\label{eq:mixed2}
\ee

 They are renormalized by the same amount through the hard modes.

Notice the $-$ sign! And that as expected the integration over the hard
modes gives a result analytic in $C$. The final result for the diagonal and off-diagonal  masses in the electrostatic
Lagrangian is obtained by adding eq.(\ref{eq:nocon}), eq. (\ref{eq:massmatrix}), eq.(\ref{eq:mixed}) and  eq.(\ref{eq:mixed2}).  The only reason the diagonal and off-diagonal masses are different is through the introduction of the constraint.

The sum of the three types of mass terms is the input in the calculation of the kinetic energy in the next section. Summaring this subsection we have for both diagonal and off-diagonal masses:
\be
m^2_E(q)={{\partial_z}^2q\over q}+O(q^2T^2)+O(g^2T^2q^2)
\label{eq:mass}
\ee
in the limit of small $q$. For the off-diagonal mass $O(q^2T^2)$ is absent, eq.(\ref{eq:massmatrix}).

For small values of $q$ we know from subsection (\ref{subsec:effectivecorrelators}) and from the explicit solution of the equation of motion for
$q$, eq.(\ref{eq:motion}), that ${\partial_z^2q\over{ q}}=m_D^2$, the physical Debye mass.

\subsection{Calculation of the kinetic energy term for small profile with ${\cal{L}}_E$}

\begin{figure} [h]
\begin{center}
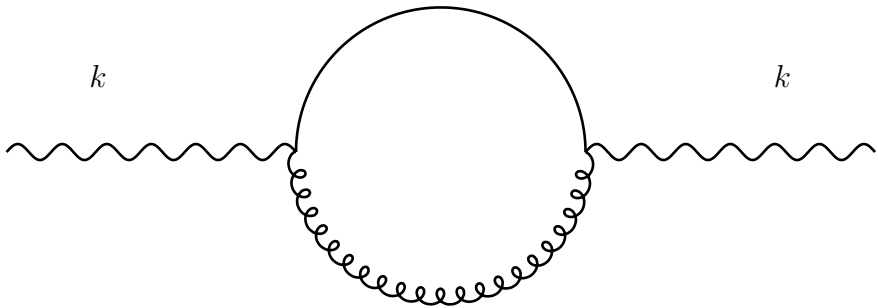
\caption{Leading contribution to the kinetic energy of the profile (wavy line). Curly line is the $Q_z$, straight line the $Q_0$ propagator. }
\label{fig:graph}
\end{center}
\end{figure}

 This has to be done with the effective Lagrangian eq.(\ref{eq:elagrangian}). The $\xi$ gauge choice is a convenient
choice because the $Q_0$ propagators decouple from the $\vec Q$ propagators.
We calculate as in the 4d case the two point function of $C_{ij}$. Linde's argument~\cite{linde} shows that with a magnetic mass $q=O(g^2)$ infinitely many diagrams contribute to order g in the kinetic energy with a coefficient $K_1$, in eq.(\ref{eq:threedimeffcontri}). There is
one diagram, shown  
in fig.(\ref{fig:graph}) that is leading by a $\log{1\over g}$ factor.
Its calculation will be described below.

 The incoming 
momentum is $\vec k=(0,0,k_z)$, the masses in the propagators are $m_E(q)$ for the off-diagonal
$Q_0$ propagator and $2\pi Tq$ for the spatial gluon propagator.
The vertices are taken from the last in term term in $V$ in the electric QCD Lagrangian. It contributes for $\xi=1$ the term proportional to the logarithm in the kinetic term\footnote{We find that the ultimate logarithmic term is $\xi$ independent} :
\be
K={(2\pi T)^2\over{2g^2}}\sum_{i\neq j}\big(1-\big(K_1g+{2g^2\over{2\pi}}{T\over{ik}}log\big({ik+2\pi qT+m_E(q)\over{-ik+2\pi qT+m_E(q)}}\big)\big)\big)
\label{eq:threedimeffcontri}
\ee

Here we took specifically $q=O(g^2)$. This is as low as we can go with $q$ in perturbation
theory. There are many multi- loop diagrams that give all $O(g)$ contributions embodied in the coefficient $K_1$ in eq.(\ref{eq:threedimeffcontri})
with this parametric form for $q$.

The result of fig. (\ref{fig:graph}) is analytic in $k=0$ and has branchcuts on the imaginary axis,
starting at $\pm i(m_E(q)+2\pi qT)$. Of course we are interested in the straight 
paths $Y_{0k}(q)$ and for $SU(2)$ there is only one, $k=1$.

 For small $q$ we have $q(z)=\exp{-m_D|z|}$, as we argued at the end of section \ref{sec:elementary}. This means
that $ik=m_D$.

The mass $m^2_E(q)$  equals $m^2_D$ plus terms of order 
$O(g^2q^2)$, according to eq.(\ref{eq:mass}).

So the onset of the branchcuts is at $m_D+2\pi Tq$, neglecting $O(g^2q^2)$.
Then  the self-energy graph, with the external legs at $m_D$, becomes 
\be
{g^2N\over{2\pi}}{T\over{m_D}}\log\big({2m_D+2\pi Tq\over{2\pi Tq}}\big)
\ee 
                                                                                
\begin{figure}[ht]                                                       \input{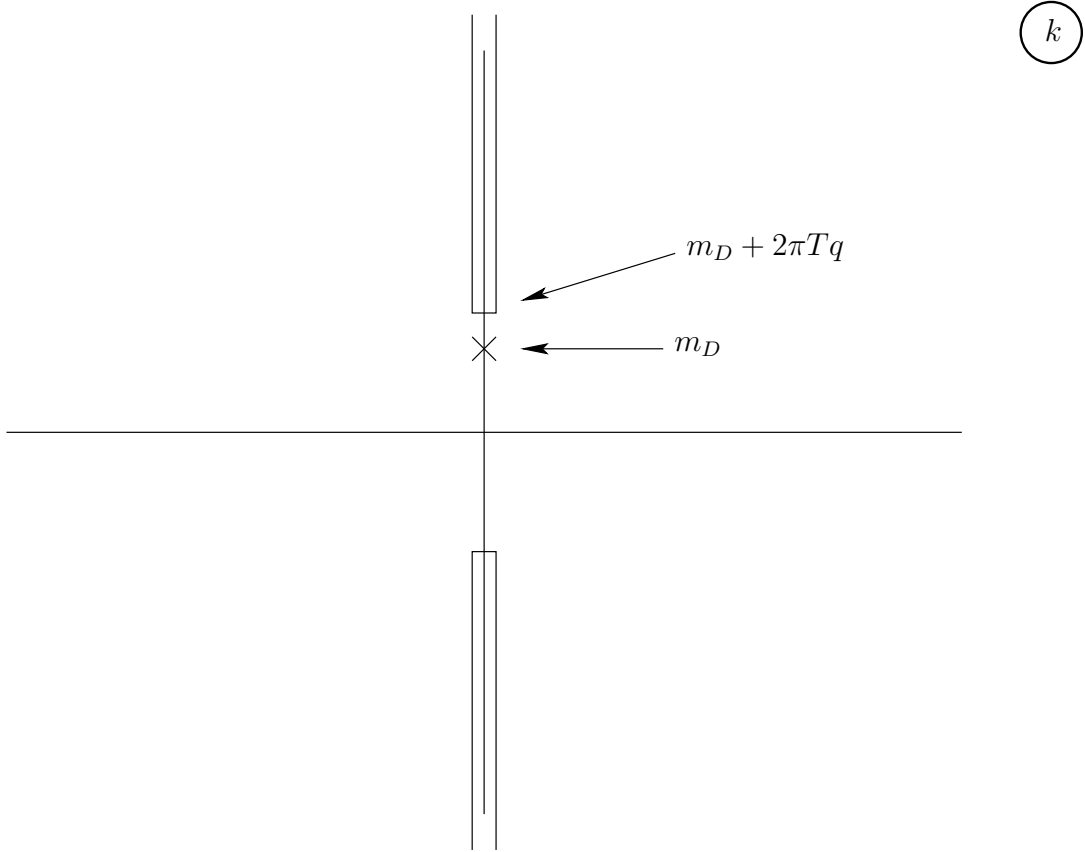}                                                    \caption[]{The complex k plane and the onset of the branchcuts for the graph in fig.(\ref{fig:graph})}           
\label{fig:plane}
\end{figure}

So the final result for the kinetic energy for small values of q is:
\be 
K={(2\pi T)^2\over{2g^2}}\big(1-(K_1g+{\sqrt{6g^2}\over{2\pi}}\log{1\over g})\big)(\partial_zq)^2
\label{eq:balance}
\ee

Our minimization equation, $K^{{1\over 2}}=V^{{1\over 2}}$, tells us that  the first term in  the kinetic energy is  balanced by the order one term $V_0$ in the potential term $V=V_0+g^2V_2$. This gives for the Debye mass the well known value $m_D=\sqrt{{2\over 3}}gT$ and that $\partial_z=O(g)$. The correction to this value comes from the second and third term in eq.(\ref{eq:balance}). But no term in $V$ can balance these terms.  So the kinetic term alone
must be zero. That is, the $O(g)$ term has to be absorbed by the Debye mass :
\be 
m_D=\sqrt{{2\over 3}}gT\big(1+g({1\over 2}K_1+ {\sqrt{6g^2}\over{4\pi}}\log{1\over g})+O(g^2)\big)
\label{eq:rebhana}
\ee
as follows from the minimization equation.

\subsection{Why there are no $O(g^3)$ contribution to the surface tension}

An immediate consequence of the last subsection was that the profile realizing the minimization equation is such that its kinetic energy to $O(g)$ is zero.

So there is no $O(g^3)$ term in the tension $2\int dq(KV)^{1\over 2}$ coming from combining the $O(g^2)$ term in $V$!
It is absorbed by the correction to the Debye mass! Eq.(\ref{eq:rebhana}) is
the well known result of Rebhan~\cite{rebhan} for $N=2$.

There is still another possibility for having an $O(g^3)$ contribution:
combining $O(1)$ in the kinetic energy with $O(g^3)$ in the potential term $V$. This can be quickly discarded. The reason is that a term $g^3V_3(q)$ is only present in a region where $q=O(g)$ or less. So integrating over $q$ in this region gives an $O(g^4)$ contribution to the tension.  

The reader may have noted that in the usual treatment of the log correction to the Debye mass the branch cut starts at $\sqrt{{2\over 3}}gT+K$, where $K$ is an ad-hoc cutoff of $O(g^2)$. The same result as in eq.(\ref{eq:rebhana}) is then obtained. But then for small enough $g$ the corrected mass will hit the branch cut! In our approach this does not happen.

Our approach can be easily generalized to the case of $SU(N)$.
The diagram  in fig.(\ref{fig:graph}) contains as incoming line the field
in one of the $Y_k$ directions. Some masses of the spatial gauge fields 
will be zero, but it is easy to see, that only the massive ones do couple,
with mass $2\pi Tq$. This is all we need to find the kinetic energy and
the log correction to the Debye mass.

\section{A numerical exploration of  the absolute minimum for SU(4)}\label{sec:absolutemin}

That the straight paths are local minima is easy to check analytically.
But for the global nature of the minima we reverted to numerical
search.

The group $SU(4)$ admits 4 minima, Two, $Y_1$ and $Y_3$, are charge conjugate. Their tension $\sigma_1$ has been computed in ~\cite{bhatta}, and is given in our eq.(~\ref{eq:tension}) with $N=4$.

\begin{figure}[ht]
\begin{center}  
\input{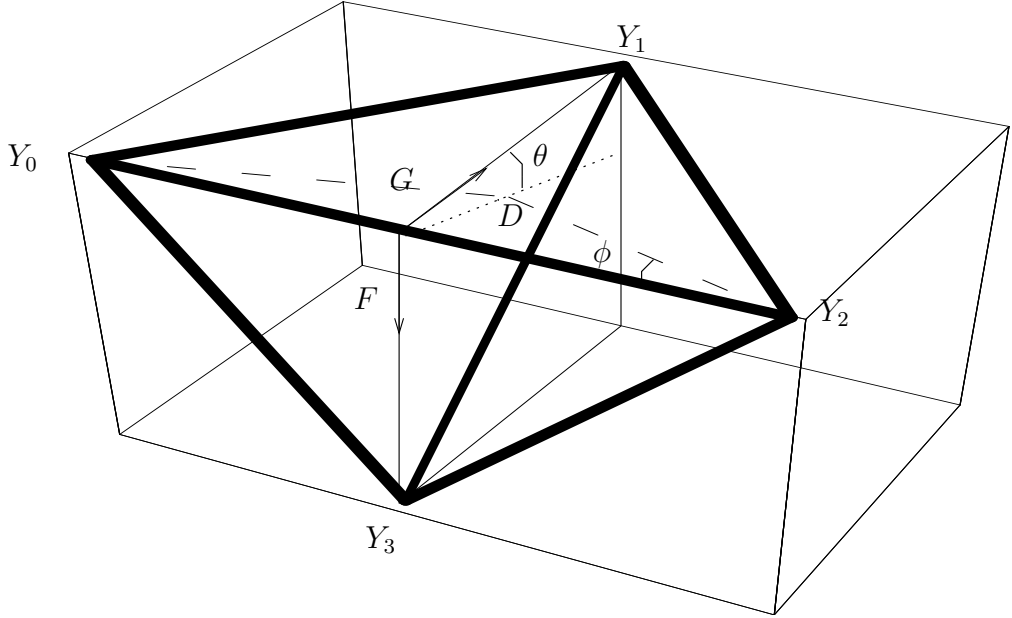}
\caption[]{Elementary cell for SU(4). The broken line connecting the corners $Y_0$ and $Y_2$ is a typical path
on which we minimized the effective action $U(C)$. Indicated are the directions of the $F$ and $G$ axes defined in the text.}
\label{fig:cell}
\end{center}
\end{figure}

So here we will be concerned with the tension $\sigma_2$ one gets by minimizing $U(C)$, eq.(\ref{eq:standardmin}) and eq.(\ref{eq:motion}), on any of the paths connecting $Y_0$ and $Y_2$ as indicated in fig.(\ref{fig:cell}). 

To see the local minimum for the path $Y_{02}(q)$ we parametrized the 3d Cartan space of $SU(4)$ with the unit  vectors ${1\over 4}Y_2$ and 
two others, $F={1\over{\sqrt 2}}(1,-1,0,0)$ and 
$G={1\over {\sqrt 2}}(0,0,1,1)$. So any point is given by
 ${q\over 4}Y_2+rF+sG$. Substitution in the potential shows that indeed a local minimum results for $r=s=0$.

To check the global minimum of the effective action minimized on the straight
path we selected a set of planes all going through the straight path $Y_{02}(q)$ and labelled by an angle $\theta$.

 Then we calculated numerically the minimal action on  a set of paths having an angle $\phi$ with the straight path from $Y_0$ to $Y_2$. So $\phi=0$ is the straight path itself.

\begin{figure}[ht]
\input{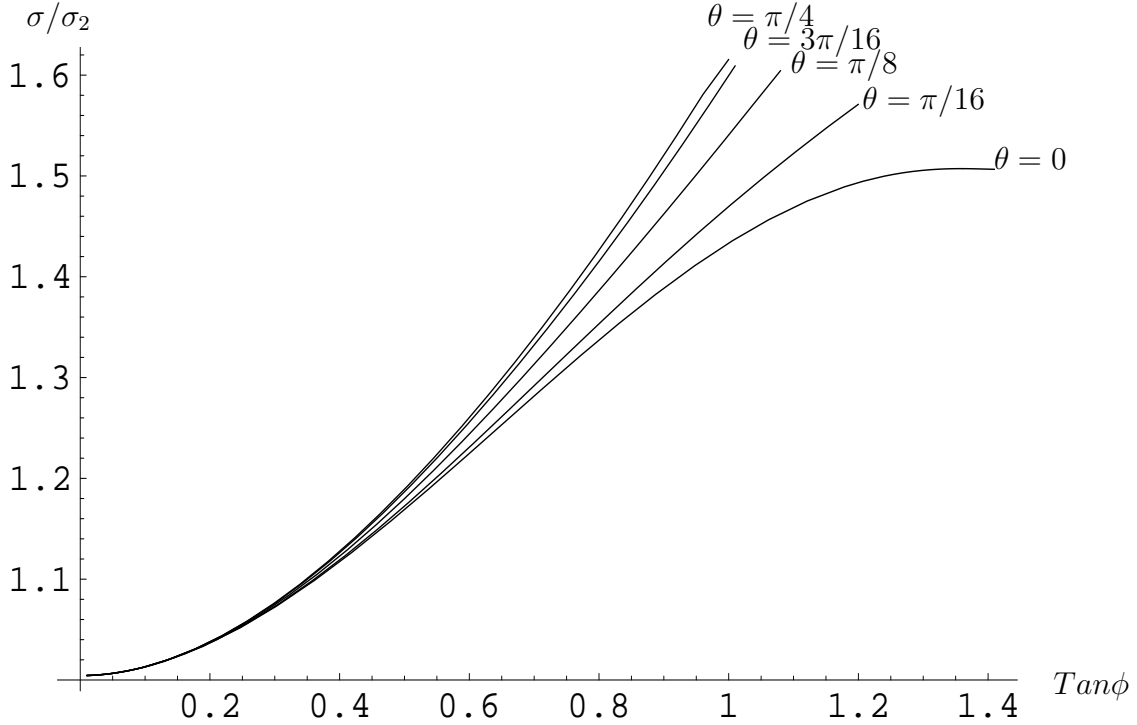}
\caption[]{The minimized action $\sigma$ on a set of paths in a given plane labeled by $\theta$. A path is selected by the angle $\phi$ as in fig.(\ref{fig:cell}). On the horizontal axis $Tan \phi$ is plotted. Endpoints of each curve are where the point $D$ hits the $Y_3Y_1$ axis in fig.(\ref{fig:cell}).}
\label{fig:theta}
\end{figure}

So in fig.(\ref{fig:theta}) each curve corresponds to a given 
plane. Because of $Z(4)$  and charge conjugation symmetry it is sufficient to look at planes    
in a quadrant $0\le\theta\le{\pi\over 4}$.

It is clear from fig.(\ref{fig:theta}), that the straight path, located at zero slope $Tan\phi=0$, and present in all
planes, so in all curves labelled by $\theta$, has globally the minimal action.

\begin{figure}
\input{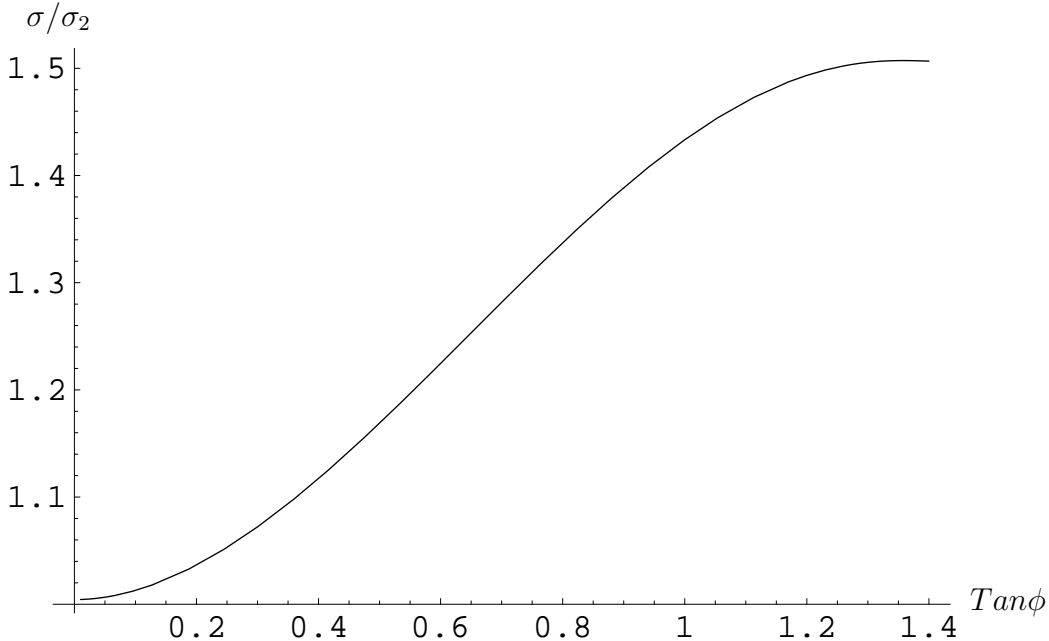}
\caption[]{The ratio for paths in the plane $\Theta=0$. The end point of the curve corresponds to $\sigma=2\sigma_1$ as explained in the text.}
\label{fig:thetazero}
\end{figure}

In fig. (\ref{fig:thetazero}) we once more plotted the result for the 
$\theta=0$ plane. Note that at the endpoint, where the path goes through
the point $Y_1$ we find $2\sigma_1$, and as expected from eq.(\ref{eq:ratio}) the value ${3\over 2}$.

\section{A simple and less simplistic approach to the ratios}\label{sec:lesssimple}

We come back to our model of gluonic screened charges we described in section (\ref{sec:elementary}).
Is it capable to reproduce the simple ratio eq.(\ref{eq:ratio})?
The first question is: what is the 't Hooft loop with k charges?
Naively one would say:

\be
V_k(L)=\exp{i{4\pi\over N}k\int d\vec S.Tr\vec E Y}
\ee

\noindent just by starting from k singly charged loops far apart and coalescing them.
However, this is wrong. It is wrong because its corresponding pathintegral expression eq.(\ref{eq:pathintegral}) tells us to go k steps in the $Y$ direction. Minimizing would then give $k\sigma_1$. But we know already that going from $C=0$ to ${2\pi\over N} TY_k$ gives us a smaller tension
 $\sigma_k=k{(N-k)\over{(N-1)}}\sigma_1$. For a concrete example (SU(4)) look
at fig.~\ref{fig:cell}. 

 The right answer is therefore:
\be
V _k(L)=\exp{i{4\pi\over N}\int_{S(L)} d\vec S.Tr\vec E Y_k}
\label{eq:rightanswer}
\ee
The reader can check that this definition satisfies the fundamental
commutation relation eq.(\ref{eq:fundamental}). When worked out in terms of
the usual pathintegral expression, as in section~\ref{sec:elementary}, it tells us to follow the minimal path from
$C=0$ to $C={2\pi\over N}TY_k$ in the elementary cell.

To see how the ratio comes about we assume that the gluons are screened, free and of course in the adjoint representation of the colour group. So we will denote the off diagonal gluons by an index pair $ij$. The diagonal gluons play no role.  

To understand the counting in the case of $V_k(L)$ we use as in subsection~\ref{subsec:simple} the adjoint Wilsonloop $TrP\exp{i\int_{L^{\prime}}d\vec l.\vec A_{adj}}$ looping $V_k(L)$.
As the reader can check with the canonical commutation relations:
\be
V_k(L)W_{adj}(L')V_k(L)^{-1}=\exp{(i{2\pi\over N}Y_{k,adj})}W_{adj}(L')
\label{eq:adjointcom}
\ee

\noindent where the hypercharge matrix in the adjoint representation is given by
its diagonal elements. These are computed from the differences of the
fundamental hypercharge matrix $Y_k$ and are therefore $0$ or $N$.
So the adjoint Wilsonloop commutes with the 't Hooft loop as it should.
But the pointlike adjoint charges send only half of the flux through
the 't Hooft loop, so produce a minus sign. And the structure of $Y_{k,adj}$ tells us precisely how to count.

Again, a gluon given by an index pair $ij$  will have 
charge $0$ or $\pm N$, with respect to the U(1) group defined by $Y_k$. Let us call this charge $l_{ij}(k)(\pm N)$. For example, if $k=1$, then all $l_{ij}(k)=0$,
except when one of the indices equals $N$, in which case $l_{ij}(1)=1$.
A gluon with generic index pair $ij$ contributes $(-)^{l_{ij}(k)}$ to the average
of $V_k$ in eq.(\ref{eq:rightanswer}).

The slab of thickness $2l_E$ around the minimal surface spanned by the loop contains   gluons, with a Poisson 
distribution $P(l_{ij})$ for each of the off-diagonal ones, $l_{ij}$
runs from $0$ to $\infty$. Their mean number present in the slab be
$\bar l_{ij}=\bar l$, assuming they are all the same. This should be the case 
because of global SU(N) symmetry. Even if the distributions were not Poissonian their widths should all be the same for the same reason. Now we  find for the average of the loop $V_k(L)$:
\bea
<V_k(L)>&=& \Pi_{ij} P(l_{ij})(-)^{l_{ij}(k)}\\        
        &=&\exp{-4k(N-k)\bar l}
\label{eq:simpleeq}
\eea
There is a factor 4 because each index pair $ij$ has a conjugate $ji$ which contributes the same to the exponent, namely $2\bar l$ or $0$.
All what goes in eq.(\ref{eq:simpleeq}) is that a gluon $ij$ with $l_{ij}(k)=0$ contributes a factor $1$ to the average, and a factor $\exp{(-2\bar l)}$ if $l_{ij}(k)=1$. The rest is counting of the index pairs that have $l_{ij}(k)=1$, and follows the reasoning just above eq.(\ref{eq:kdepv}).

A non-Poissonian width will contribute a common ($ij$ independent)  factor in front of the exponent.

The result for the ratio eq. (\ref{eq:ratio}) follows immediately, since $\bar l=n(T)Al_E$ where
$n(T)\equiv n_{ij}(T)$ is the density of a charged gluon $ij$.

So also the ratio is a consequence of the gluons being screened, free and
in the adjoint multiplet of $SU(N)$.

\section{Spatial Wilsonloops}\label{sec:wilson}

About the Wilsonloops there is no analytic knowledge, unlike 
the 't Hooft loops. What we know is from dimensional arguments, and from 
simulations~\cite{karschtension}.

The dimensional argument says that very high temperatures the physics of
our system is determined by 3d gauge theory with coupling $g_3^2=g^2T$.
So the surface tension $\rho$ will be proportional to this mass scale squared:
\be
\rho\sim g^4T^2
\label{eq:wilsondim}
\ee

Another important fact from simulations is that in the hadronic phase the 
surface tension is constant from $T=0$ to $T=T_c$. This is consistent with
the idea that in this phase the cause of the area law is colour magnetic flux in the $T=0$
groundstate, and that the hadronic gas at $T\neq 0$ does not contribute any 
such flux.

Above the critical temperature, according to the dimensional argument and 
corroborated by simulations~\cite{karschtension}, the string tension behaves as
if the plasma phase {\it{itself}} provides colour magnetic flux.

So the question becomes: what is providing this flux?

There are basically two answers: the first is macroscopically
long flux loops. Their presence in the plasma is unlikely since
it is difficult to excite them thermally. Only will those loops
contribute  that  maybe already present in the groundstate of cold QCD.
So they will not cause the {\it{rise}} in the Wilson surface tension with temperature.

The second option is monopoles. Their fluxes are screened in the plasma.
 The Dirac monopoles with flux ${2\pi\over N}$
are too singular. 
 
Then there are 't Hooft-Polyakov type monopoles with screened flux. In what follows we will concentrate on them. What we have to say here is speculative.   

We will assume there are screened free monopoles at very high T in the plasma. They are supposed to have charge ${2\pi N\over g}$. An old argument
by Goddard et al.~\cite{goddard} suggests that the magnetic group is
given by $SU(N)/Z(N)$ if the gauge group is $SU(N)$, or $SU(N)$ if the gauge group is $SU(N)/Z(N)$. We suppose the monopoles to be in the adjoint representation. 
They are consistent with the presence of free quarks with charge ${g\over N}$. For $N=2$ they have the same charge as  the 't Hooft-Polyakov monopoles.

With this hypothesis we can deduce the area law for the Wilson loop, and 
the ratios for Wilsonloops with centergroup charge $k$. Not surprisingly 
it will turn out to be identical to that of the 't Hooft loops, eq.(\ref{eq:ratio}). Not surprising because our input is an adjoint multiplet of monopoles
dual to the multiplet of gluons. And the counting of the effective gluon
charges, as in eq.(\ref{eq:adjointcom}), has a dual as we will see shortly.

We start with the area law in $SU(2)$.

We need a representation of the Wilson loop that clearly shows how it captures the flux of an 't Hooft-Polyakov monopole.
 
In a recent  paper by one of the authors~\cite{korthalskovner2} the following representation of the 
spatial Wilsonloop in pure SU(2) was found by precisely a procedure of 
continuation of the 't Hooft-Polyakov monopole flux through a loop L from the Higgs phase to the symmetric phase (the VEV of the Higgs goes to zero) and then letting the Higgs mass go to infinity, which is essential to give
the expression a well-defined meaning. What one finds is that the fundamental Wilsonloop equals:
\bea
W_1(L)&=&\int \Pi_{\sigma,\tau,a} Dn^a(\sigma,\tau)\exp{{i\over 2}\int_{S(L)} d\vec S.(\vec B^an^a+\epsilon^{abc} n^a\vec Dn^b\wedge \vec Dn^c}) \\
      &=&\int D\Omega\exp{i\int_LdsTr{\tau_3\over 2}\big(\Omega A_s(s)\Omega^{-1}+{1\over i}\Omega\partial_s\Omega\big)}
\label{eq:wilsonloop}
\eea
The unit vector $\vec n$ is defined as:
\be
\vec n=Tr\Omega^{-1}\tau_3\Omega{\vec\tau\over 2}
\ee

The integration over $\vec n$ shows the fluctuations over the directions of the Higgs field that come in naturally when letting the VEV of the Higgs go to
zero.
These representations were first derived by Diakonov and 
Petrov~\cite{diakonov}. They prove extremely useful in the present context. 
In the first representation we have a clear connection with the integrated 
flux of an 't Hooft-Polyakov monopole. That of  the second equation is  clearly a rewriting of
the Wilson loop. The path integral in the first expression is over the unit vector  $\vec n$
defined on any choice of surface with as border $L$. The second path integral 
is over all SU(2) gauge transformations periodic on the loop $L$.

In order to compute the average of the Wilsonloop 
let us assume that our monopoles, being screened, are free. This makes only sense if their mean distance exceeds the screening length $l_M$~\footnote{See the discussion section for discussion of this point}.

 With one monopole present within a distance $l_M$ of the loop the first representation gives a 
factor $-1$ for the loop.

 How does this happen? 
Like for the electric charges in the presence of the 
't Hooft loop the monopole is carrying a magnetic charge vis \`a vis the $\tau_3$.
So the monopole is off-diagonal and picks up the difference of the eigenvalues
of $\tau_3$. Note the factor ${1\over 2}$ in front of the exponent. The outcome 
is therefore $\pi$ for the flux in the exponent.

Then with the now 
familiar Poisson distribution $P(l)$ for the monopoles in the slab of thickness
$2l_M$ we find the area law:
\be
<W_1(L)>=\sum_l(-)^lP(l)=\exp{-2\bar l}
\ee
and the mean number of monopoles in the slab is $\bar l=2l_MAn_M(T)$.

As mentioned in the introduction the density of monopoles is therefore related 
to the surface tension of the Wilson loop and using eq. (\ref{eq:wilsondim}):
\be  
n_M(T)\sim g^6T^3
\ee
So the density of the monopoles compared to that of the gluons is down by a factor
$({1\over {\log({T\over{\mu_T}})}})^3$.

There is an important consistency check on our assumption that the 
monopoles are free and screened. As we said before, their density should be small enough
with respect to the screening length. So we should compute the number of monopoles within the screening radius, $l^3_Mn_M(T)$. From the relation between the tension and the density it follows:
\be
l^3_Mn_M(T)\sim l^2_M\rho(T)
\label{eq:consistent}
\ee
Only for SU(2) both the screening mass~\footnote{Obtained in ref. (\cite{rebbi}) for a lattice of temporal extent 4 at $\beta=2.82$.} from the correlation of heavy Dirac monopoles  and the tension~\cite{karschtensionsu2} are numerically available:
\bea
l^{-1}_M&=&0.86g^2(T)T\\
\rho(T)&=&(0.369\pm 0.014)^2g^4T^2
\eea

So we find for $SU(2)$ that the number of monopoles inside the screening radius is $\sim 0.19$, or the mean distance between two monopoles is
about twice the screening length~\footnote{We neglected in the data a  discrepancy in subtraction points. At large enough $T$ this is irrelevant.}.  

Finally the pressure will get a positive contribution to the coefficient
of  $O(g^6)$.

\subsection{Ratios of Wilson surface tensions}

Let us turn to the ratios of surface tensions for SU(N), with $N>3$. For that we need  a formula
for the multiply charged Wilsonloop $W_k(L)$. We take again the matrix $Y_k$ and
write:
\bea
W_k(L)&=&\int \Pi_{\sigma,\tau,a}Dn^a_k(\sigma,\tau)\exp{{i\over N}\int d\vec S.\big(\vec B^an^a_k+f^{abc}n^a_k\vec Dn_k^b\wedge \vec Dn_k^c\big)}\\
 &=&\int D\Omega(s)\exp{{i\over N}\int ds TrY_k\big(\Omega(s)A_s(s)\Omega^{-1}(s)+{1\over i}\Omega(s)\partial_s\Omega^{-1}(s)\big)}
\label{eq:wk}
\eea

The field $n^a$ describes the coset $SU(N)/SU(k)\hbox{x}SU(N-k)\hbox{x}U(1)$
which has $2k(N-k)$ dimensions.
As in the case of the 't Hooft loop it is defined by
\be
n^a_k=TrY_k\Omega{\lambda^a\over 2}\Omega^{-1}
\ee
where $\Omega$ is any regular $SU(N)$ gaugetransformation.

The reader can verify that the fundamental commutation relations ~\ref{eq:fundamental} are satisfied, using the identity $TrY_kY_l=N^2\hbox{min}(k,l)-klN$.

The length of the vector $n^a_k(\vec x)$ does not depend on $\vec x$:
\be
\sum_an^a_kn^a_k={1\over 2} TrY_k^2
\ee

The counting of the monopoles that do contribute to the tension $\rho_k$
goes in full analogy with that under eq.(\ref{eq:adjointcom}). The difference is that 
we  now ask what the commutation relation is between an 't Hooft loop in the adjoint representation
$V_{adj}(L)$ and the Wilsonloop $W_k(L')$:
\be
V_{adj}(L)W_k(L')V^{-1}_{adj}(L)= \exp{i{2\pi\over N}Y_{k,adj}}W_k(L')
\ee
with $Y_{k,adj}$ the $Y_k$ charge in the adjoint representation.

So the loops do commute. But now the pointlike monopoles send only 
a flux ${2\pi N\over 2}$ through the Wilson loop, so will contribute $-1$.

It is now familiar from section~\ref{sec:lesssimple} how to proceed with the counting of monopoles that do indeed contribute $-1$. An off-diagonal monopole $ij$ will contribute
 only when $ij$ is one of the $2k(N-k)$ combinations admitted
by $Y_k$. 
The Wilson surface tension becomes proportional to:
\be
\rho_k\sim k(N-k)\bar l
\ee

\noindent where $\bar l$ is the average number of $ij$ type monopoles in the slab of width $2l_M$, taken to be the same for all.

We then find the scaling for the ratio:
\be
\rho_k={k(N-k)\over{N-1}}\rho_1
\ee

\section{Discussion}

 In this paper we have looked at the physics of the plasma through the 't
Hooft loop and the Wilson loop. The former is on a firm basis, the latter of a more speculative nature, but both give testable results. For the former we showed:\\

\noindent i) that the surface tension associated to the 't Hooft loop has a calculable
perturbative expansion\\

\noindent ii)that it obeys, including two loop order and up to order $O(g^4)$ a  simple dependence on the $Z(N)$ charge in Cartan space
of the loop  minimized along the straight path\\

\noindent iii)that the Debye mass absorbs all $g^3$ effects of the surface tension.\\

\noindent iv) that the Feynman rules coming from the constrained pathintegral have a natural infrared cut-off built in.\\

  For the lowest order the  result ii) was evident, for the two loop result it is due to the factorization into two Bernoulli functions.  For three loops the factorization is gone, so one has to check
by different methods wether the simple dependence still holds.
We should point out that there are statements in the litterature~\cite{smilga}
that already the second order contribution has non perturbative contributions.
That this is not the case is demonstrated in section~\ref{sec:twoloopratio}.

Let us remember that this same simple dependence came out of the plasma 
viewed as a gas  of free  screened gluons. From that point of view  the
interactions could easily change the behaviour.

For SU(4) our result for the 't Hooft loop is on a firm basis. We checked that the straight path is the global minimum. What is painfully lacking is some analytic understanding of this fact in SU(4), and  for higher N. What singles out the straight paths for all values of $N$ is that the Wilson line effective potential renormalizes multiplicatively along these paths, something that was known up to now only for the path from $0$ to $Y_1$ giving the singly
charged 't Hooft loop tension.
The ratios at the critical point (which we could obviously not compute with the techniques at hand) will give valuable information on the 3D spin model describing the universality class of the transition.
 
Point iv) suggest doing multi-loop calculations for the free energy with these Feyman rules, in particular the perturbative logarithm at sixth order.

Let us now discuss the more speculative aspects for the Wilson loops.

For the Wilson loop we could understand on the basis of lattice results that at high $T$ a second component of  screened and free monopoles
can explain the tension and the numbers available for $SU(2)$ suggested this is a consistent picture. 

The magnetic contribution to the sixth order pressure has now a definite sign and is related to the surface tension of the Wilsonloop measured by Karsch~\cite{karschtension} through $l_M$,  a quantity measured from the correlation of the Dirac monopole loops~\cite{rebbi}.

 The reader might be shocked by the statement that within the screening radius there are few
monopoles. For the statistical Debye screening for the gluons  one has {\it of necessity}
many gluons within the screening radius $l_E$: $l_E^3n_E(T)\sim g^{-3}$.

On the other hand the magnetic screening has {\it{not}} only statistical origins,
because also at low temperatures it is still present~\cite{rebbi}.
So  we do not see any flagrant inconsistency
in the assumption that there is a dilute gas of screened monopoles present in the plasma at very high $T$. It is dilute  because it only has to explain the ``small''  Wilson loop tension $\rho(T)\sim g^4T^2$. On the other hand the density
of gluons has to explain the ``large'' semiclassical result for the 't Hooft loop tension, $\sigma(T)\sim{1\over g}T^3$.

If it is indeed the dominant contribution to the sixth order term in the pressure, it would because of its positive sign push the pressure up from the known fifth order, and nearer 
to the Stefan-Boltzman limit. We do {\it {not}} pretend that it sheds light
on the   perturbative convergence of the pressure in  the plasma phase of SU(N) gauge theory! In fact a  recent letter~\cite{recentkaj} to understand the convergence needs a {\it negative} sixth order coefficient. This result is {\it not} in contradiction with ours! The sixth order coefficient consists of the 
non perturbative contribution and a perturbative one. The latter consists of a term
coming from the $T$ scale, and a log term with argument $l_D/l_M$. Knowledge of the
perturbative part would clearly be helpful.

For the ratios of Wilson loops we speculated that these monopoles were in the adjoint representation of a magnetic group ($SU(N)$ or $SU(N)/Z(N)$).
Then the same argument that worked for gluons gave the dependence on the 
centergroup charge $k$ of the Wilsonloop.

In the ratios the unknown constant from the statistical distribution  drops out and they are the same as those for the 't Hooft loops.

In order to get a more quantitative grip one should have a precise lattice 
 measurement of  $l_M$, the sixth order coefficient of the pressure, the perturbative logarithmic term therein, and the surface tension of the spatial Wilsonloop. The $SU(2)$  measurement of the magnetic screening mass should be done for $SU(3)$, so that together with the spatial Wilson loop results one can check
the consistency of the monopole model in that case as well.

A measurement in $SU(4)$ and $SU(5)$ of ratios at high $T$ would settle some of the issues. 
 In the case of the spatial Wilson tension at very high $T$ we have the option of
simulating in the $T=0$ 3D theory~\cite{karschlaer}. 

A remark applying to our results for both types of loops: their surface tensions do not depend on the representation, only on the N-allity. This follows from our discussion of their effect on physical states in subsection~\ref{subsec:simple}. The surface effects can {\it{only}} come from the 't Hooft commutation relations. The latter are determined by the N-allity
of the loops. As expected, Teper and Lucini~\cite{teperlucini} find the same for temporal Wilson loops in $T=0$ $SU(4)$
and $SU(5)$ theory. This issue is still debated though~\cite{bali}. 
  
What happens when dynamical quarks are present in our plasma? As we noted,
they are consistent with the monopole component through the Dirac condition. So the Wilson loops will not be affected by their presence. But the 
't Hooft loops will be affected. They will depend on the surface $S(L)$,
not anymore only on  the loop $L$. These effects can and should be studied. So our plasma would have three screened components at high temperatures: gluons,
quarks and monopoles. The latter are by far the largest screened objects,
and in density down by a factor $(log{T\over {\Lambda_T}})^{-3}$. At lower $T$ they start to coalesce. 

The results described in ref.~\cite{digiacomo} have bearing on the magnetic
activity in the cold phase and its behaviour in the hot phase, i.e. on the 
temperature independent part of the Wilson loop. Our monopoles have to do with
the rise of the tension of the Wilsonlop. The fact that our monopoles are screened may be due to this magnetic activity.

As a last question: what
happens to the ratio in 3d gauge theories? There the answer is: the same
as in 4d gauge theories for the analoque of the 't Hooft loop. That is: take the correlation between two vortices with 
strength $k$ and compare it to that with strength $k=1$. Both have stringy behaviour and
the ratio of the stringtensions is as in eq.(\ref{eq:ratio}).

 For the Wilson loops the situation is different. The tension is caused by the 't Hooft $Z(N)$
vortices. If one assumes that  vortices with different charge have the same density at high enough $T$ we find the ratio is one, but we expect in general a dependence.

When finishing this paper we were informed about the existence of string/brane inspired computations of ratios of Wilson loops at zero temperature~\cite{strassler}. There the ratio is in the  notation adopted here:
\be
{\rho_k\over {\rho_1}}={\sin{{\pi\over N}k}\over{\sin{{\pi\over N}}}}
\ee

This has in common the linearity in ${k\over N}$ present in our result.
This result is consistent with recent findings on the lattice
~\cite{teperlucini}.
Of course the physics in both cases is very different so it is not clear 
wether comparison makes much sense.

\section*{ Acknowledgements}

We thank  Marc Knecht, S. Bronoff and R. Buffa for helpful discussions
on the infrared behaviour of the kinetic term. Input by R. Buffa was 
important. Correspondence with Robert Pisarski
,  John Wheater and Juerg Froelich
on universality were quite informative. Philippe de Forcrand shared many of his insights with us. Mikko Laine provided us with his understanding of the perturbative
approach in hot QCD.  Useful discussions with Mike Teper, Biagio Lucini, Alex Kovner and Ian Kogan were indispensable. Correspondence with Christian Hoelbling
is gratefully acknowledged. Incisive remarks by Pierre van Baal and Jan Smit were very helpful.

 The funding of the Royal Society and the CNRS for the bilateral Oxford-Marseille cooperation is gratefully acknowledged. 
 One of us (P.G.) thanks the MENESR for funding.

\section*{Appendix A. Bernoulli polynomials}
The hatted Bernoulli polynomials appearing in the text are related to the Bernoulli polynomials in Ryzhik-Gradshtein, as given below.

$$\widehat B_4(x)={2\pi^2\over 3}T^4\left(B_4(x)+{1\over 30}\right)
\eqno A.1$$
$$\widehat B_3(x)={2\pi\over 3}T^3B_3(x)\eqno A.2$$
$$\widehat B_2(x)={T^2\over 2}B_2(x)\eqno A.3$$
$$\widehat B_1(x)=-{T\over 4\pi}B_1(x)\eqno A.4$$

with
$$B_4(x)=x^2(1-\vert x\vert)^2$$
$$B_3(x)=x^3-{3\over 2}x^2\epsilon(x)+{1\over 2}x$$
$$B_2(x)=x^2-\vert x\vert$$
$$B_1(x)=x-{1\over 2}\epsilon(x)$$

All $B$'s are defined mod 1. If the subscript is even, the $B$ is even; if odd the $B$ is odd. So in particular $B_1$ is a sawtooth function.


\section*{References}
\begin{thebibliography}{99}

\bibitem{thooft}G.'t Hooft, {\it Nucl. Phys.} {\bf B138}, 1 (1978)

\bibitem{korthalskovner}C.P. Korthals  Altes, A. Kovner and M. Stephanov, {\it Phys. Lett.} {\bf B469}, 205 (1999), hep-ph/9909516. 

\bibitem{bhatta} T. Bhattacharya, A. Gocksch, C. P. Korthals Altes and R. D. Pisarski, 
{\it Phys. Rev. Lett.} {\bf 66}, 998 (1991);
{\it Nucl. Phys.} {\bf B 383} (1992), 497.

\bibitem{svetitski} B. Svetitski, L.G. Yaffe, Nucl. Phys. B210, 423 (1980).

\bibitem{linde} A. D. Linde, Phys. Lett. B96, 289 (1980).

\bibitem{iancu} J.P. Blaizot, E. Iancu, A. Rebhan, Phys. Rev. Lett 83 (1999) 2906, hep-ph/9906340; J.O. Andersen, E. Braaten, M. Strickland, Phys. Rev. Lett. 83 (1999), 2139, hep-ph/9902327.

\bibitem{bronoff}S. Bronoff, R. Buffa, C.P. Korthals Altes, hep-ph/9801333. 

\bibitem{deforcrand}Ph.de Forcrand, M. D'Elia, M. Pepe, hep-lat/0007034. 

\bibitem{mandelstam} G. 't Hooft, in High Energy Physics, ed. A. Zichichi(Editrice Compositori Bologna, 1976); S. Mandelstam, Phys. Rep. 23C,(1976) 245. 
 
\bibitem{digiacomo} A. di Giacomo, B. Lucini, L. Montesi, G. Paffuti, Phys. Rev. D61, 034503 (2000); H. Shiba, T. Suzuki, Phys. Lett. B333, 461 (1994);
M. Chernodub, M.I. Polikarpov, A Veselov, Phys. Lett. B342, 303 (1995).

\bibitem{korthalsaltes}C. P. Korthals Altes, \Journal{\NPB}{420}{637}{1994} 

\bibitem{rebbi} Ch. Hoebling, C. Rebbi, V. A. Rubakov, hep-lat/0003010 v2.

\bibitem{wingate} S. Ohta, M. Wingate, hep-lat/0006016.


\bibitem{weiss} N. Weiss {\it Phys. Rev.}{\bf D24} (1981) 475; {\bf D25} 
(1982) 2667. 

\bibitem{dvali} S. Bronoff, G. Dvali, K. Farakos, C. P. Korthals Altes, Proceedings Strong and Electro Weak Matter 1997, eds Z. Fodor, A. Patkos , World Scientific, 1998.

\bibitem{huanglissia}S. Huang, M. Lissia, Nucl. Phys. B438, 54,1995 
, hep-ph/9411293. 

\bibitem{braatennieto}K. Farakos, K. Kajantie, K. Rummukainen, M. Shaposhnikov NPB425(1994)67, 
hep-ph/9404201; E. Braaten, J.A. Nieto, Phys. Rev. Lett. 76, 1417 (1996).

\bibitem{giovan}P. Giovannangeli, C.P. Korthals Altes, in preparation.

\bibitem{belyaev}V. M. Belyaev, Phys. Lett. B254 (1991),153.

\bibitem{kovnerfossil}A. Kovner, hep-ph/0102329.

\bibitem{arnoldyaffe}P. Arnold, L. Yaffe, Phys. Rev. D52,7208 (1995).

\bibitem{teperhart}A. Hart, M. Teper, Phys. Rev. D60 (1999) 114506, hep-lat/9902031.


\bibitem{kajantie} K. Kajantie, M. Laine, J. Peisa, K. Rummukainen, M. Shaposhnikov, Phys. Rev. Lett. 79, 3130 (1997), hep-ph/9708207.

\bibitem{karschtension} F. Karsch, E. Laermann, M. Lutgemeier,
 Phys. Lett. B346,94 (1995), hep-lat/9411020. 

\bibitem{rebhan}A.K. Rebhan, \Journal{\PRD}{48}{3967}{1993}.

\bibitem{gross}Mark Gross, John Bartholomew, David Hochberg,  EFI-83-35-CHICAGO; J.F. Wheater, Mark Gross, Phys. Lett. B144, 409 (1984). 

\bibitem{recentkaj} K. Kajantie, M. Laine, K. Rummukainen and Y. Schroeder, Phys. Rev. Lett.86, 10, 2001, hep-ph/0007109.

\bibitem{korthalskovner2} C. P. Korthals Altes, A. Kovner, Phys. Rev. D62,
096008, hep-ph/0004052.

\bibitem{pisarski}P. Ginsparg, Nucl. Phys. B170, 388 (1980); T. Applequist, R. D. Pisarski, Phys. Rev. D23, 2305 (1981), S. Nadkarni, 
Phys. Rev. D27, 917 (1983).

\bibitem{karschtensionsu2}G.S. Bali, J. Fingberg, U.M. Heller, F. Karsch, K. Schilling, Int. J. Mod. Phys. C4 (1993), 1179, hep-lat/9308003.

\bibitem{diakonov}D. Diakonov, V. Petrov, Phys. Lett. B224, 131 (1989).

\bibitem{goddard}P. Goddard, J. Nuyts, D. Olive, Nucl. Phys. B125, 1 (1977).

\bibitem{karsch}F. Karsch, M. Lutgemeier, A. Patkos, J. Rank, Phys. Lett. B390, 275(1997), hep-lat/9605031.

\bibitem{teperlucini}M. Teper, B. Lucini, hep-lat/0012025.

\bibitem{bali}G.S. Bali,  Phys. Rev. D62, 114503 (2000), hep-lat/0006022. 


\bibitem{karschlaer}F. Karsch, E. Laermann and M. L\"utgemeier, Phys. Lett. B346 (1995) 94, hep-lat/9411020.

\bibitem{smilga}A.V. Smilga, Phys. Rep. 291, 1 (1997).

\bibitem{strassler}M.J. Strassler, Nucl. Phys. Proc. Suppl. 73 (1999) 120.
\end{thebibliography}
\end{document}